\documentclass[]{aa}
\usepackage{natbib}
\usepackage{graphicx}
\usepackage{subfigure}
\usepackage{captcont}

\begin{document}

\title{Three clusters of the SMC from ACS/WFC HST archive data:
          NGC~265, K~29 and NGC~290 and their fields
population }
\subtitle{}

\author{Emanuela Chiosi\inst{1}, Antonella Vallenari \inst{2}}

   \institute{Astronomy Department, Padova University, Vicolo
          dell'Osservatorio 2, 35122 Padova, Italy \\
   \and   INAF, Padova Observatory,
Vicolo dell'Osservatorio 5, 35122 Padova, Italy     \\
   \email{emanuela.chiosi@oapd.inaf.it; antonella.vallenari@oapd.inaf.it  }
             }

\offprints{A. Vallenari}

\date{Received: December 2006 / Accepted: January 2007 }

\abstract{}{We determine the age, metallicity and initial mass
function  of three clusters, namely \object{NGC~265}, \object{K~29}, \object{NGC~290}, located
in the main body of the Small Magellanic Cloud. In addition, we
derive the history of star formation in the companion fields.}
 {We
make use of ACS/WFC HST archive data. For the clusters, the age and
metallicity are derived fitting the integrated luminosity function
with single synthetic stellar population by means of the $\chi^2$
minimization. For the companion fields,  the history of star
formation is derived using the  $\chi^2$ minimization together with
the downhill-simplex method.} 
{For the clusters we find the
following ages and metallicities:  \object{NGC~265}
  has    log(Age)=$8.5\pm0.3$ yr and
metallicity  $0.004\pm0.003$(or [Fe/H]=-0.62); \object{K~29} has   log(Age)=$8.2\pm0.2$ yr
and  metallicity Z=$0.003\pm0.002$ (or [Fe/H]=-0.75); \object{NGC~290} has 
log(Age)=$7.8\pm0.5$ yr and  metallicity  $0.003\pm0.002$(or [Fe/H]=-0.75). The
superior quality of the data allows the study of the initial mass
function down to M$ \sim$ 0.7 M$_\odot$. The initial mass function
turns out to be in agreement with the standard Kroupa model. The
comparison of the  \object{NGC~265} luminosity function with the theoretical
ones from stellar models both taking overshoot from  the convective
core into account and neglecting it, seems to suggest that a certain
amount of convective overshoot is required. However, this conclusion
is not a strong one because this cluster has a certain amount of
mass segregation which makes it  difficult to choose  a suitable area
for this comparison. The star formation rate of the field population
presents periods of enhancements at 300-400 Myr, 3-4 Gyr and finally
6 Gyr. However it is relatively quiescent at ages older than 6 Gyr.
This result suggests that at older ages, the tidal interaction
between the Magellanic Clouds and the Milky Way was not able to
trigger significant star formation events.  } {} 

\titlerunning{Three clusters of the SMC}
\authorrunning{E. Chiosi \& A. Vallenari}

\maketitle

\section{Introduction}\label{sec_intro}

It is well known that the rich cluster system of the Magellanic
Clouds (MCs) is quite different from that of our own Galaxy where a
dichotomy is found between young sparse open clusters and old
compact objects. The MCs present a wide distribution of rich cluster
ages and constitute an ideal laboratory for testing  the star and
cluster formation process, either due  to global effects, such as
gravitational triggers because of tidal interactions,  or to
internal processes. In fact, the star formation (SF) process depends
on the cooling and the heating of the interstellar gas which in
turn, depend on the presence of metals
\citep{lisenfeld1998,wolfire1995}. The Large and Small Magellanic
Clouds (LMC and \object{SMC}, respectively)  are particularly important as
they allow us to study the SF in low metal content environments. A
large population of clusters is found in the \object{SMC}. \citet{hodge1986},
comparing the number  of clusters found down to $B$=22 (inner
regions) or $B$=23 (outer regions) in selected regions with the
number of known clusters in SMC catalogs at that time, estimated a
global population of 900 clusters, but more than 2000 clusters are
expected   if small-size, older clusters were detectable.

Concerning the cluster age distribution, several catalogs are
available in literature.  \citet{pietrzynski1999} using isochrone
fitting, and \citet{rafelski2005} making use  of integrated colors
derive the age of a limited number of bright clusters, namely 93,
and 200 respectively. \citet{chiosi2006} derive the age of about 164
associations and 300 clusters using isochrone fitting on OGLE II
data. However only a few of the clusters in those catalogs have ages
derived from high quality CMD data. Ground based CMDs  must avoid
the cluster centers and suffer from the effects of the high crowding
and serious contamination by field stars. The superior quality of
Hubble Space Telescope (HST) data favors the study of the central
regions of the clusters allowing precise determinations of their
ages. Among the few age determinations of SMC clusters based on HST
data, we   quote
%
%
\citet{rich2000} who presented HST/WFPC2 data of 7 clusters.

 In this paper, as a part of a
project aimed to cast light on the process of field and cluster
formation in the MCs, we present HST ACS/WFC archive data of three
SMC clusters, namely \object{NGC~265}, \object{K~29} and \object{NGC~290} observed by Olszewski
in 2004 (Proposal ID=10126) as part of the ACS/WFC program, for which no high quality
CMDs have been obtained up to now. They are located close to the
borders of HI shell  37A where star formation seems to have been
very active in the recent past \citep{stanimirovic2004, chiosi2006}.
The coordinates of the clusters are given in Table \ref{coord.tab}.
Preliminary age determinations on the basis of OGLE II data and
isochrone fitting can be found in \citet{chiosi2006} who, assuming
the typical value for the metallicity (Z=0.004) and the reddening
appropriate to each clusters, estimated log(Age) = 8.4, 8.0, 7.8 yr
for \object{NGC~265}, \object{K~29}, and \object{NGC~290}, respectively.

The aim of this paper is to derive better estimates of the age and
metallicity of the three clusters,  to check the slope of their
initial mass functions (IMF) down to $M \sim 0.7M_\odot$, and to
derive the star formation history (SFH) of the companion field
populations.

In section \S \ref{obs} we present  the data and describe the
reduction procedure. In section \S \ref{modulus} we discuss the SMC
distance modulus, the reddening and the metallicity. In section \S
\ref{age_det} we describe  the techniques used to derive age, metal
content, and IMF. In section \S \ref{param} we derive the physical
parameters of the clusters.  In section \S \ref{fields} we determine
the SFH of the nearby fields. Finally, in section \S \ref{conclusion}
we draw the final remarks.

\begin{table}
\label{coord.tab} \caption{Equatorial coordinates of the SMC regions
containing the three clusters.}
\begin{center}
\begin{tabular}{ l c c }
\hline
Cluster  &   RA (J2000) &  DEC (J2000)\\
\hline
\object{NGC~265}   &   0:47:12  &   -73:28:38\\
\object{K~29}      &   0:51:53  &   -72:57:14\\
\object{NGC~290}   &   0:51:14  &   -73:09:41\\
\hline
\end{tabular}
\end{center}
\end{table}

\section{Data  presentation,  reduction and calibration}\label{obs}

{\bf Reduction}. We make use of HST archive data taken in 2004 using
the Advanced Camera Survey (ACS/WFC) in the filters  F555W and
F814W. Two exposure times per filter are available, namely 120s and
440s. We make use of pre-reduced data that are already corrected for
the geometric distortion of the camera. The data reduction is made
with the DAOPHOT II/ALLSTAR packages  by \citet{stetson1994}. In
brief, the CCD images are split in two regions, one containing the
cluster and the other  the field, the aperture photometry is
calculated with an aperture of 3 pixels, the point spread function
(PSF) is  derived for each colour and each CCD, and the PSF
photometry is performed. Data are finally corrected for the charge
transfer efficiency (CTE) using the \citet{riess2004}
transformations. This correction goes from 0.01 mag for the
brightest magnitudes to 0.1 mag for the faintest magnitudes.



{\bf Calibration}.
%
The transformation from the instrumental magnitudes to the
ACS/WFC-Vega system, taken from \citet{bedin2005}, is given by

\begin{eqnarray}
 m_{filter}   &\equiv &  -2.5 log_{10} \frac{I_{e^-}}{exptime} +
 Z_p^{filter}  \nonumber \\
 & &  -\Delta m _{PSF-AP(r)}^{filter}-\Delta m _{AP(r)-AP(\infty)}^{filter}
 \end{eqnarray}

\noindent where $exptime$ is the exposure time, $Z_p^{F555}=25.718$
and $Z_p^{F814}=25.492$ \citep{bedin2005},  and $\Delta m
_{PSF-AP(r)}^{filter}$ and $\Delta m _{AP(r)-AP(\infty)}^{filter}$
are the two terms of the aperture correction. The first one compares
the PSF magnitudes with the aperture magnitudes at a finite
magnitude (3 pixels in our case) and the second one is the
difference between aperture magnitude at finite size and  magnitude
at
 infinite aperture.
The first term, which is  tabulated in Table \ref{calib.tab}
separately for the two parts in which each CCD frame has been split,
is statistically calculated for the brightest magnitudes as the mean
difference.
 We assume as aperture
corrections from r=3 pixel to infinity the value given by
\citet{sirianni2005}: $\Delta m _{AP(r)-AP(\infty)}=0.243$ for the
F555W filter,  $\Delta m _{AP(r)-AP(\infty)}=0.291$ for the F814W
filter.

\begin{table}
\caption{Aperture corrections for both filters F555W and F814W. CCDs
are split in 2 parts: the upper row (1) refers to the cluster
whereas the lower row (2) to the field.} \label{calib.tab}
\begin{center}
\begin{tabular}{ l l | c | c }
\hline
\\
\multicolumn{2}{l}{Cluster} &
\multicolumn{1}{l}{$\Delta m
_{PSF-AP(3pix)}^{F555W}$} &
\multicolumn{1}{l}{$\Delta m _{PSF-AP(3pix)}^{F814W}$} \\
\\
\hline
\object{NGC~265} & (1)   & -0.14 $\pm$ 0.02   & -0.18 $\pm$ 0.04 \\
        & (2)   & -0.15 $\pm$ 0.02   & -0.21 $\pm$ 0.03 \\
\object{K~29}    & (1)   & -0.14 $\pm$ 0.02   & -0.19 $\pm$ 0.02 \\
        & (2)   & -0.15 $\pm$ 0.02   & -0.19 $\pm$ 0.03 \\
\object{NGC~290} & (1)   & -0.14 $\pm$ 0.02   & -0.18 $\pm$ 0.02 \\
        & (2)   & -0.13 $\pm$ 0.02   & -0.18 $\pm$ 0.03 \\
\hline
\end{tabular}
\end{center}
\end{table}

ACS/WFC magnitudes are converted to the Johnson-Cousins  V, I following
\citet{sirianni2005}. The magnitudes are compared with V,I ground-based
data
%
%
from OGLE II \citep{Udalski1998} survey for \object{NGC~265}, \object{K~29} and
\object{NGC~290}. We make use of the stars brighter than V=18 and fainter
than V=17.3 to avoid  deviation from linearity due to saturation on
HST data.
%
%
%
The mean magnitude differences  are $\Delta_V=
V(HST)-V(OGLEII)=0.02\pm0.04$ and $\Delta_I=
I(HST)-I(OGLEII)=0.01\pm0.04$. These small differences allow us to
take the magnitudes of the saturated stars in the ACS photometry
from the OGLE II catalog (see the sections below for more detail).
%

%

\begin{figure}
\centering \resizebox{\hsize}{!}{\includegraphics{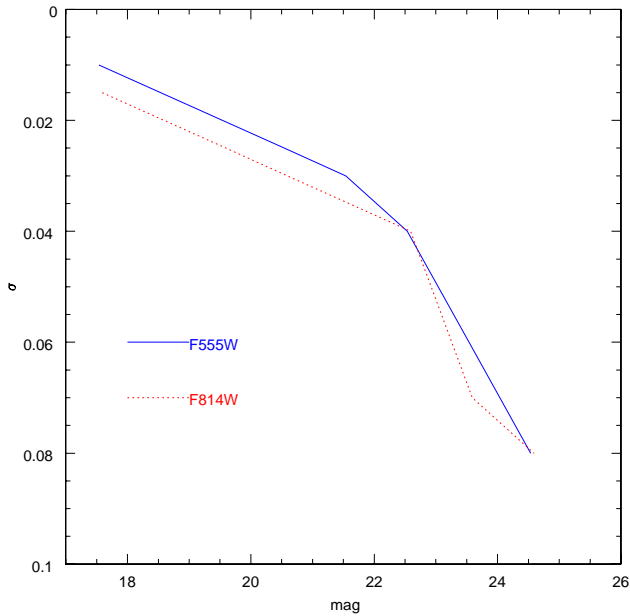}}
\caption{Mean photometric errors $\sigma$ as function of the  magnitude for the F555W and
F814W pass-bands for  the studied fields.}
\label{errors}
\end{figure}

We estimate the photometric errors in the two pass-bands by means of
the artificial stars experiments as the difference between the
assigned and recovered magnitude for any simulated star. The results
are shown in Fig. \ref{errors} as a mean of all the studied fields.
This is justified by the fact that differences from field to field
are not relevant.

The  observational CMDs of the three clusters are presented in the
top left panels of  the Figs. \ref{ngc265_clu.fig},
\ref{k29_clu.fig}, and \ref{ngc290_clu.fig}. They  will be examined
in  detail below. They refer to a  circular area having 22--25$"$ of radius
(depending on the cluster)
inside which the vast majority of the cluster stars falls as shown
by the surface brightness profiles  in the top right panels of the
Figs. \ref{ngc265_clu.fig}, \ref{k29_clu.fig}, and
\ref{ngc290_clu.fig} (see the discussion below).

{\bf Incompleteness}. Correction for incompleteness is derived as
usual by means of artificial star experiments where artificial stars
of known magnitude are added to the original image and data are
re-reduced. The ratio between the number of  recovered stars and the
original number of added stars gives the completeness correction
$\Lambda$. The results are shown in the bottom right panels of the
Figs. \ref{ngc265_clu.fig}, \ref{k29_clu.fig}, and
\ref{ngc290_clu.fig} which present $\Lambda$ as function of the
magnitude for the inner radius of the  clusters.
%
%
and for the  external regions. More details
can be found in the  sections below.
%


\section{Distance modulus, reddening and metallicity}\label{modulus}

The distance modulus  $(m-M)_0=18.9$  is assumed for the SMC, in
agreement with recent determinations  by \citet{storm2004},
\citet{weldrake2004}, \citet{caputo1999} and \citet{sandage1999}.\\

No assumptions are made for the reddening and metallicity of each
cluster. They are derived together with the age  from the  analysis
of the CMD  by means of the $\chi^2$ minimization technique to be
discussed below.

In the study of the field population,  assumptions have to be made
for the SMC depth along the line of sight, interstellar extinction
and age-metallicity relation. We proceed as follows:

\noindent (1) We adopt the SMC depth along the line of sight  of 4
kpc, which implies  a difference in the distance modulus of about
0.14 mag \citep{welch1987,zaritsky2002}.

\noindent (2) The extinction map across the SMC  derived by
\citet{zaritsky2002} shows that  the interstellar reddening is
spatially varying, i.e. increasing along the SMC ridge from
North-East to South-West.
 A mean reddening $E(B-V)$=0.08 mag is
derived by \citet{tumlinson2002} and \citet{hunter2003}.
 However,  we prefer to use for the field population the same extinction
 we have obtained for the study of the  CMD of the companion cluster using
 the minimization procedure.
%

\noindent (3) Our  knowledge of the age-metallicity relation in the
SMC is mainly based on  clusters.  The interpretation of the
existing SMC age-metallicity relation widely varies from author to
author. It goes  from  continuous enrichment from the oldest to the
youngest objects as found  by \citet{dacosta1998} and by
\citet{dolphin2000b} to a burst-mode of star formation and
enrichment as derived by \citet{olszewski1996}, \citet{pagel1999},
\citet{piatti2001}, \citet{harris2004}. To study the star formation
history in the fields, we assume the enrichment history by
\citet{pagel1999}.

\section{Method to derive ages, metallicities, and IMFs}\label{age_det}

 Age, metal content, and IMF are estimated by comparing theoretical
luminosity functions (LF) that depend on the three parameters under
examination to the observational one and by plotting the
corresponding isochrones on the CMD to check the mutual consistency
of the results. The determination is first made by visual inspection
of the CMD  and it is  refined  with the aid of the $\chi^2$
minimization. Theoretical luminosity functions and isochrones are
taken from the library of \citet{girardi2002}.

Corrections for completeness and field subtraction are the
preliminary step to be undertaken to derive good experimental
luminosity functions. The analysis articulates as follows:

\noindent 1. Equal areas are considered for each cluster and
companion field.

\noindent 2. The luminosity interval spanned by the cluster and
field stars in each pass-band is divided in number of intervals of
0.5 mag width. In each interval, and separately for the cluster and
field, we count the stars and apply the correction for incompleteness
(the cluster and field completeness factors are used as appropriate)
and finally we calculate the ratio

 \begin{equation}
  N_{true}=\frac{1}{\Lambda} \times N_{obs}
 \end{equation}

\noindent where $\Lambda$ is the completeness factor.

\noindent 3. In each magnitude interval the  field stars are
statistically subtracted from the cluster stars.

\noindent 4. The above procedure is applied to derive both the LF
and the CMD of the cluster. The corrected LFs for each cluster are
presented in  Figs.
\ref{ngc265f_lum_cfr1.fig}, \ref{k29pop_isto.fig}, and
\ref{ngc290pop_isto.fig} together with their theoretical
counterparts.

\noindent 5. From the  \citet{girardi2002} isochrone library we
select one of suitable age and metallicity that best matches the CMD
 paying particular care to the
location of the main sequence,
 and  to the luminosity of the core He-burning stars.
%
%
Once  the above estimates of age and metal content are obtained we
refine them by means of the  $\chi^2$ minimization of the LF. We
start from the previous estimates, use the Padova simulator of
synthetic CMDs, that can take  the photometric errors on magnitudes
into account, and apply the Monte Carlo technique to generate many
simulations of the LF and companion CMD at varying age, metallicity,
reddening,  and slope of the IMF. Finally the $\chi^2$ minimization
technique is applied to pin down the best combination of the
parameters.

The IMF is derived in the mass range from  the turnoff mass to  the limit set
by the completeness of the photometry, i.e. at about  $0.7
M_{\odot}$. We assume an IMF of the form:

\begin{equation}
dN \propto M^{-\alpha}dM
\end{equation}

\noindent where the slope $\alpha$ is a free parameter and is
derived in two mass ranges, namely  $0.7M_{\odot}<M<1M_{\odot}$ and
$1M_{\odot} <M<4M_{\odot}$. In the adopted notation, the
\citet{1955ApJ...121..161S} slope is $\alpha =2.35$.

{\bf Binary stars}. A large fraction of stars in cluster and field
CMDs are unresolved binaries, either physical binaries or chance
superposition of stars along the line of sight.   Their presence
blurs the main sequence, the giant branch, the clump of red stars,
and the loop, thus making it more difficult to derive precise
estimates of age, metallicity and IMF slope. Particularly cumbersome
is the use of the turn-off or termination magnitude of the main
sequence to infer the age by fitting isochrones and luminosity
functions when binaries are present.
 The effect of
unresolved binary stars on the CMD and luminosity functions of
stellar clusters has been investigated in many  studies among which
we recall \citet{C89a}, \citet{V91}, and \citet{Barmina}. As long
known, the presence of unresolved binaries would first brighten the
turn-off and termination magnitude and broaden the main sequence.
Second if the photometry is particularly good they would even split
the main sequence into two parallel loci, the one of unresolved
binaries being systematically cooler and brighter (about 0.7 mag for
binary stars of equal mass) than that of single stars, thus
facilitating the task of assigning the true age and metallicity. Our
present data do  not allow this. Therefore we are left with the
brightening of the turn-off and termination magnitude which would
mimic a younger age and also yield a different slope for the IMF in
the upper mass range. For the three clusters under examination we do
not have information on their percentage of binary stars  nor on the
distribution of the mass ratio and separation of these latter.
Despite this, we apply the synthetic CMD technique to include the
effect of unresolved binaries.  We assume that binary stars obey the
same IMF of single stars and the mass ratio falls in the range $0.7$
to $1$. Systems with mass ratios different from these cannot be
excluded. However they would be hardly distinguishable from single
objects. Finally, we suppose that the percentage of unresolved
binaries is the same as indicated by  open clusters of the Milky Way
that  amounts to about 30\,\% -- 50\,\% of the cluster population
\citep{Mermi-Mayor89,Ca94}, and other well studied clusters of the
LMC such as NGC1818 \citep{El98} and  NGC1866 \citep{Barmina}. Here
we adopt the percentage of 30\%.

\begin{figure*}[t]
\parbox{8.7cm}{
\resizebox{8.7cm}{!}{\includegraphics{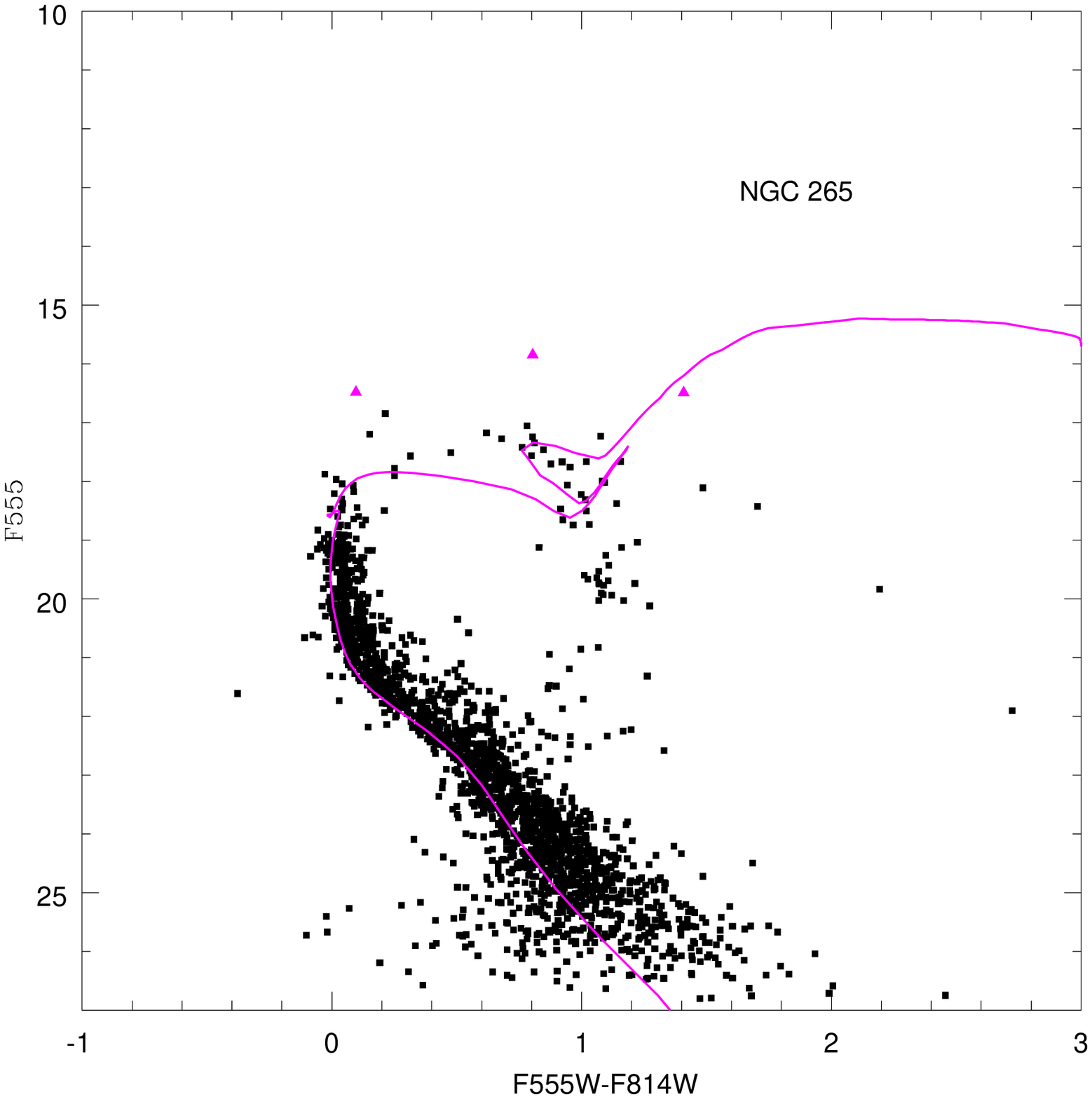}}  \\
}\hspace{0.2cm}
\parbox{8.7cm}{
\resizebox{8.7cm}{!}{\includegraphics{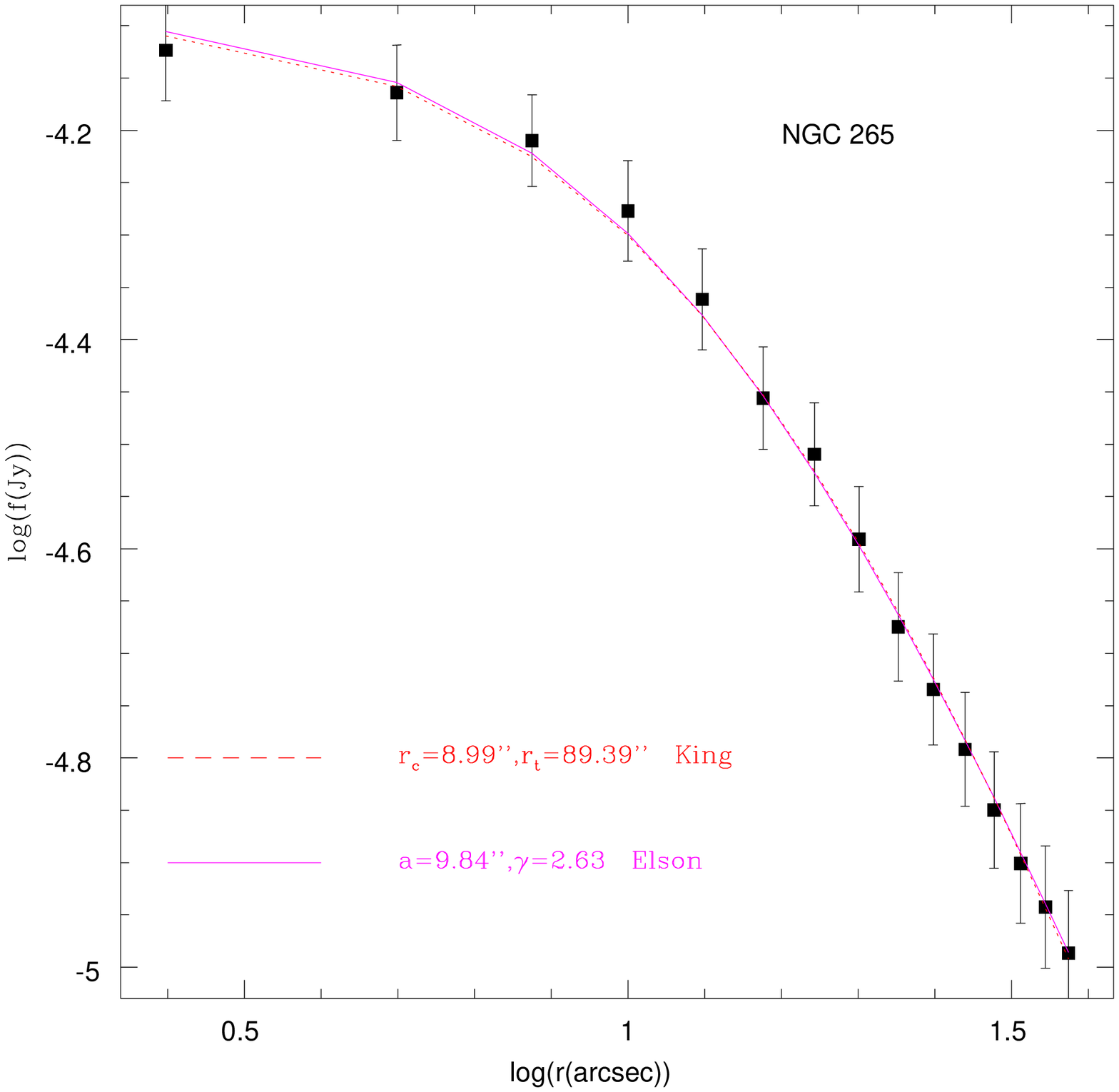}}\\
}\hspace{0.2cm}
\parbox{8.7cm}{
\resizebox{8.7cm}{!}{\includegraphics{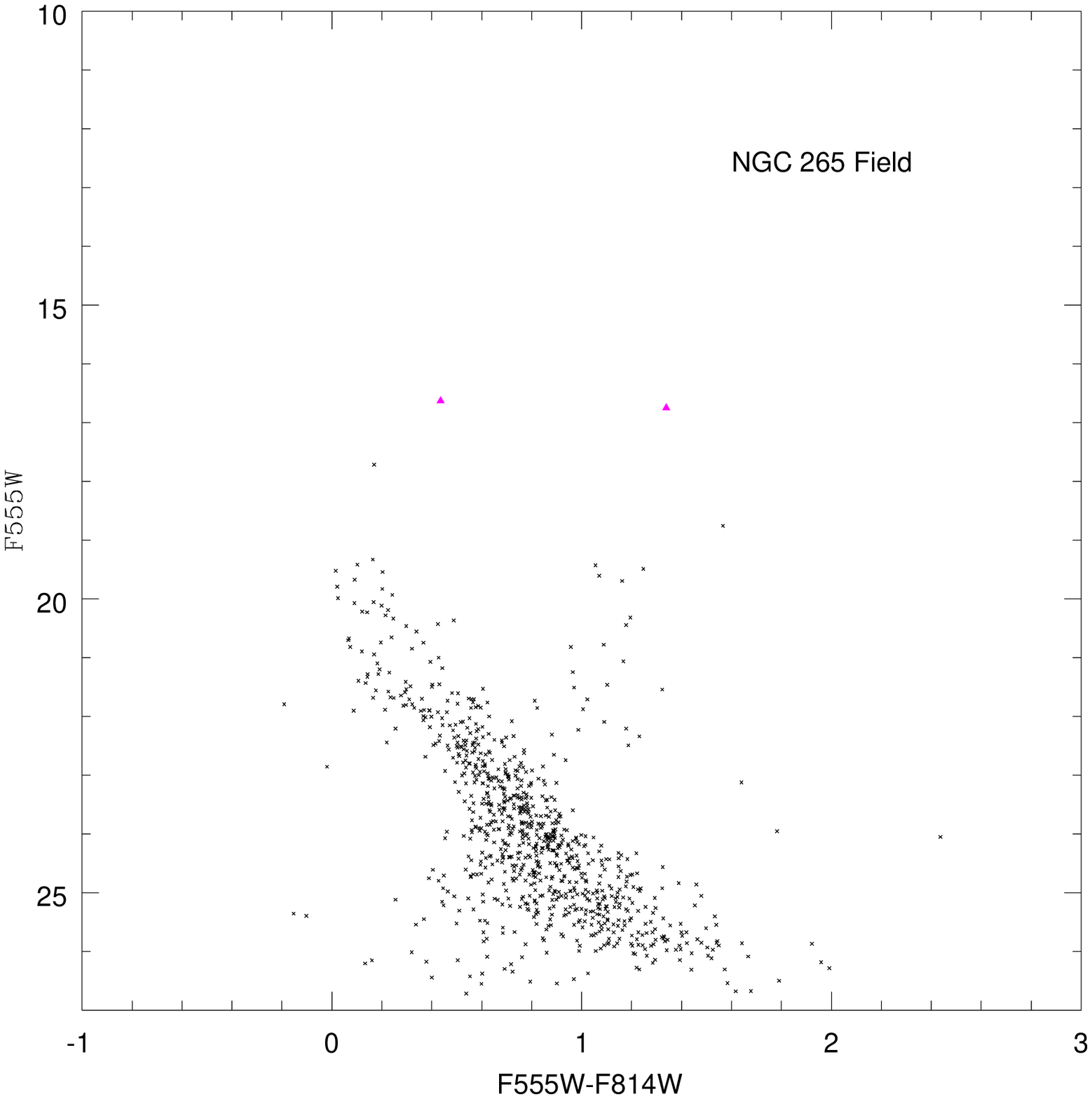}}\\
}\hspace{0.2cm}
\parbox{8.7cm}{
\resizebox{8.7cm}{!}{\includegraphics{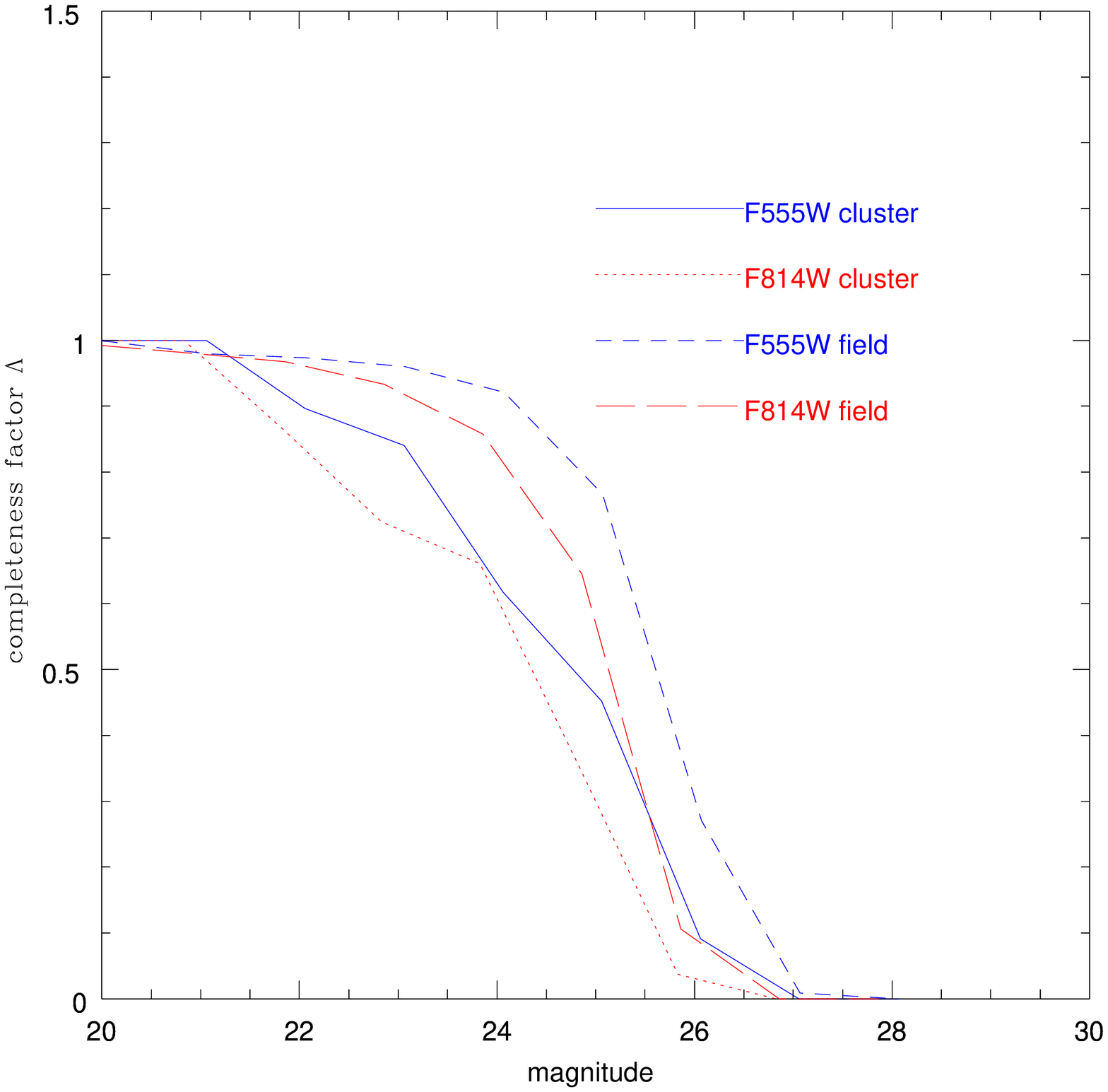}}\\
}
 \caption{Summary panels for  \object{NGC~265}. In the top left  panel is the
CM diagram (inner 22'') with a superposed isochrone of log(age)=8.5,
Z=0.004 and E(F555W-F814W)=0.08. The brightest stars settle at
F555W=16. In the top right panel the surface brightness profile (in
arbitrary units) is  fitted by one of the King functions (dashed
line) and by  Elson profile (continuous line). From this
distribution we can determine core and tidal radius.
 Vast majority of the stars sets within a radius of 22''
 corresponding to log(r)=1.3. In
the bottom left panel the CMD of field stars  is shown. The field
population is taken at about 1.6' from the cluster center and the
area is comparable to  the area of the cluster (see text for
details). The bottom right panel shows the  completeness factors for
the  cluster and field regions as indicated} \label{ngc265_clu.fig}
\end{figure*}

\begin{figure*}[t]
   \parbox{8.7cm}{
   \resizebox{8.7cm}{!}{\includegraphics{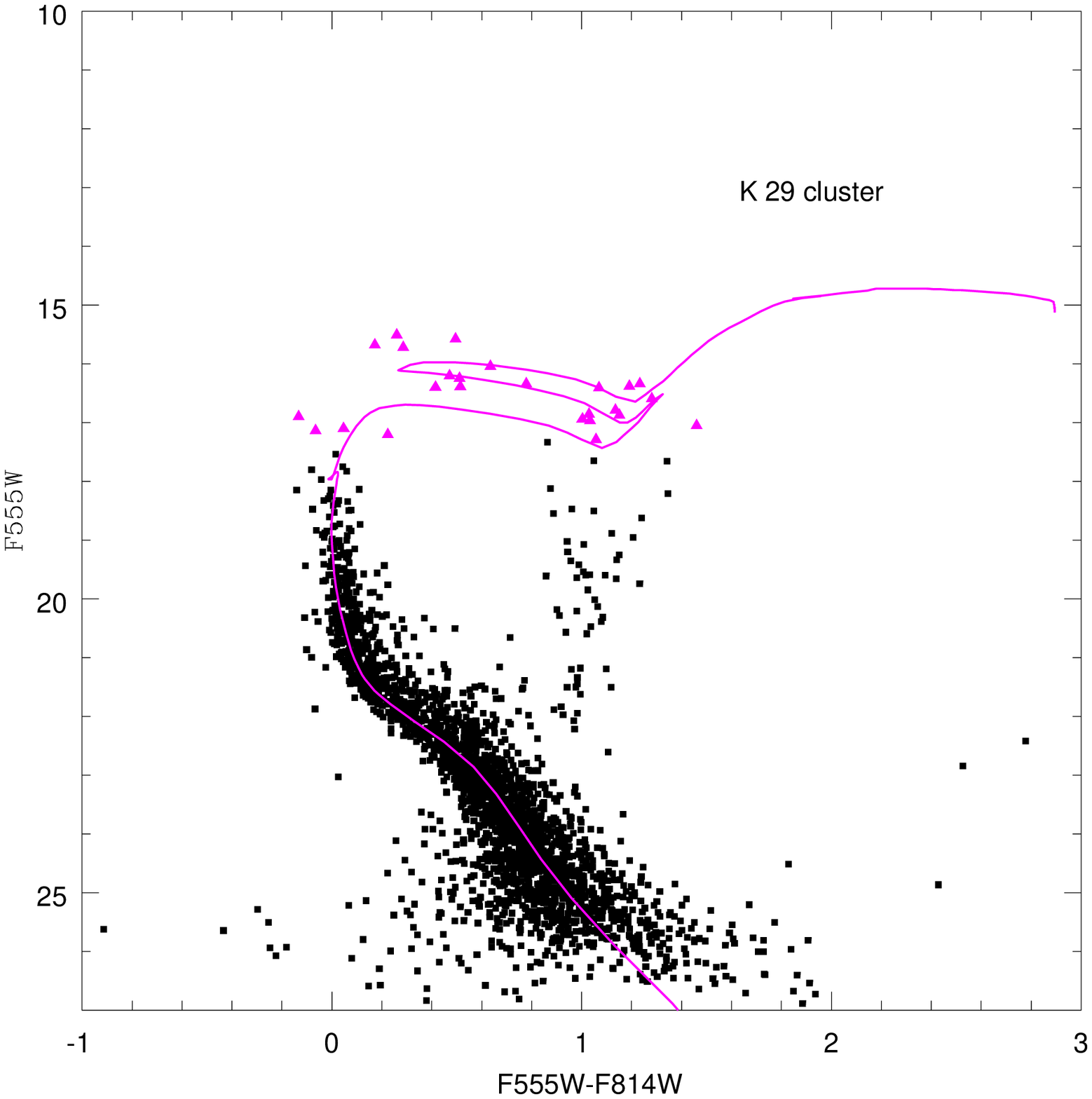}} \\
     }
  \hspace{0.2cm}
  \parbox{8.7cm}{
   \resizebox{8.7cm}{!}{\includegraphics{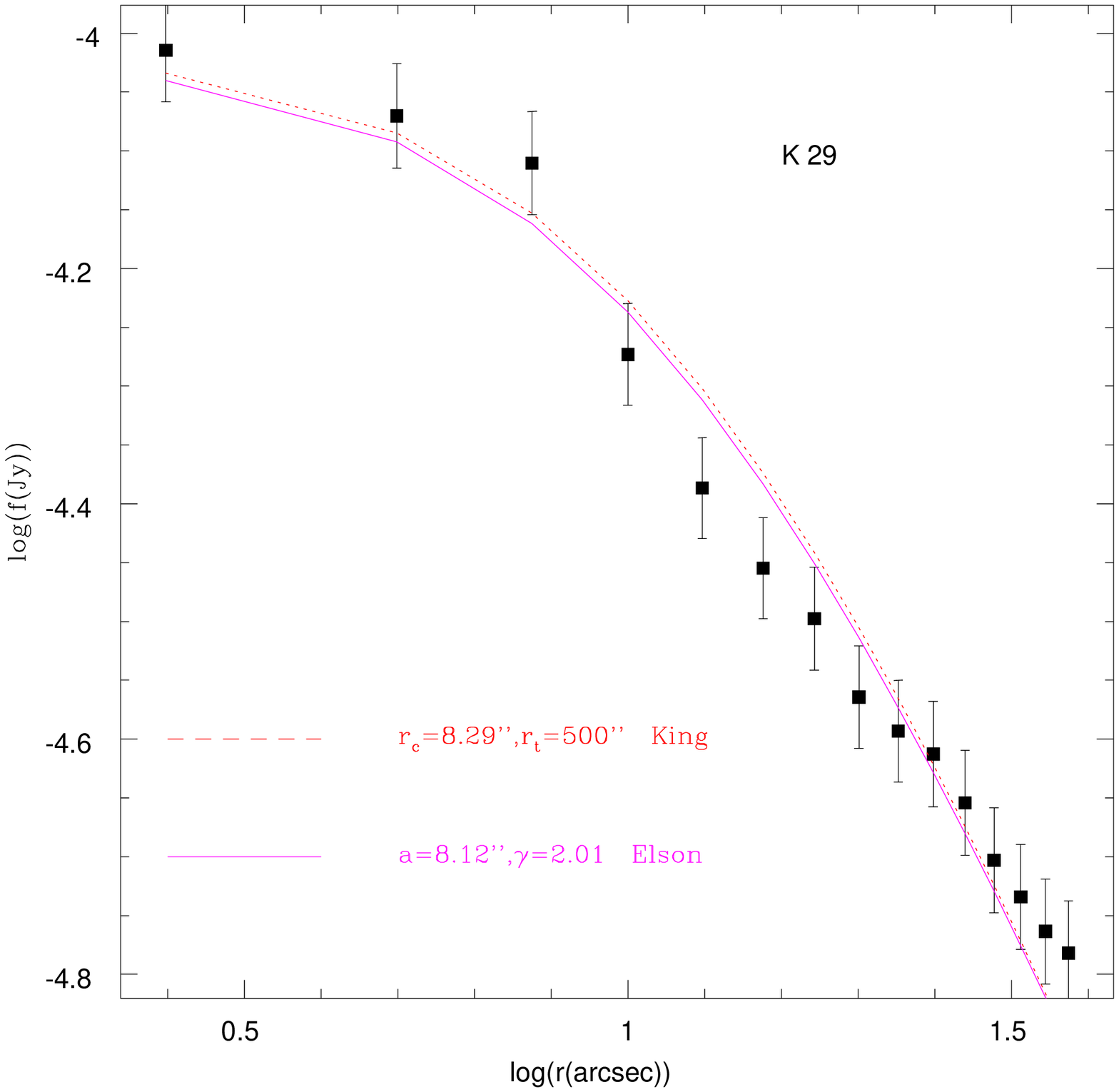}}\\
   }
   \hspace{0.2cm}
  \parbox{8.7cm}{
   \resizebox{8.7cm}{!}{\includegraphics{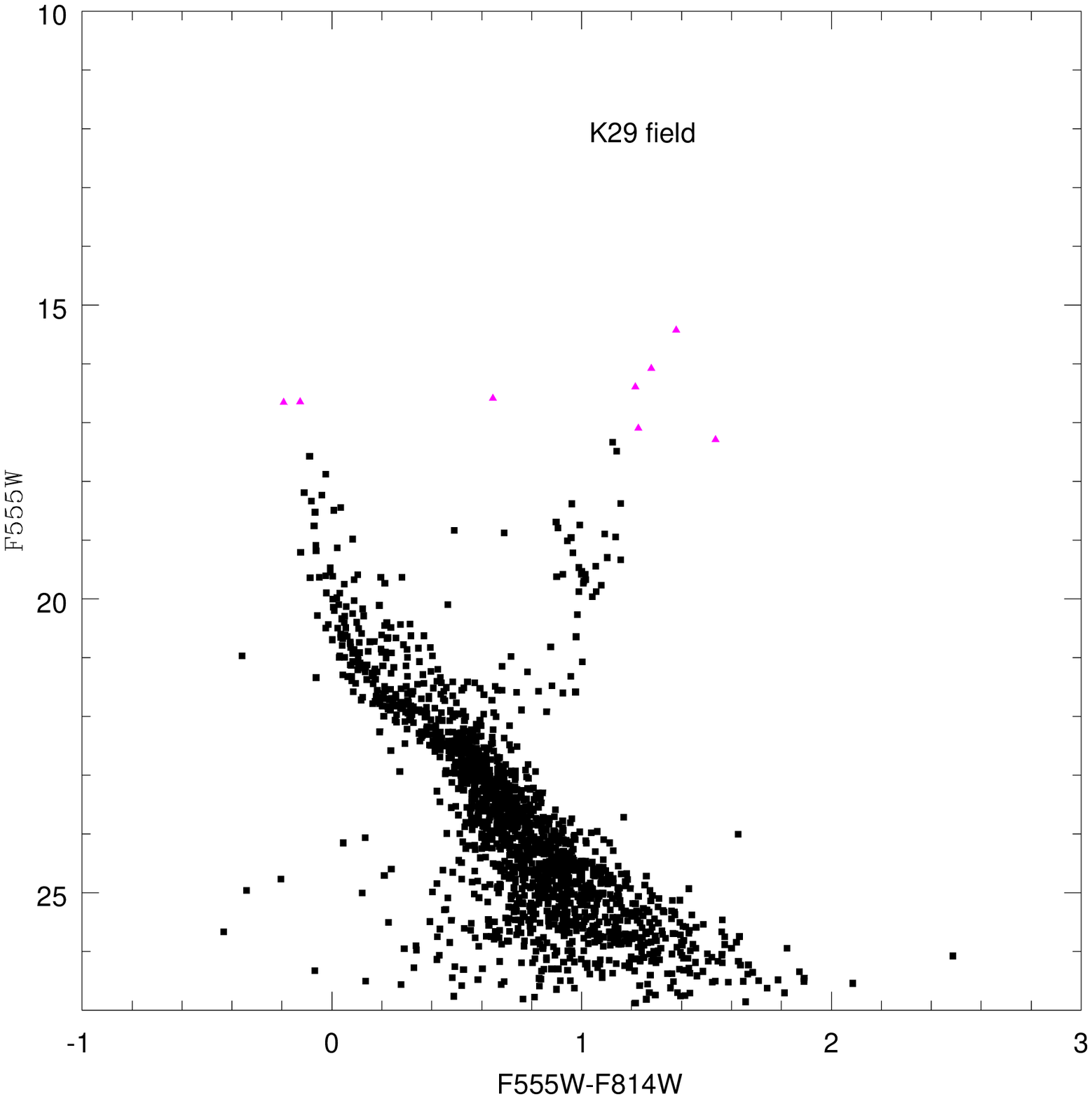}}\\
   }
  \hspace{0.2cm}
  \parbox{8.7cm}{
   \resizebox{8.7cm}{!}{\includegraphics{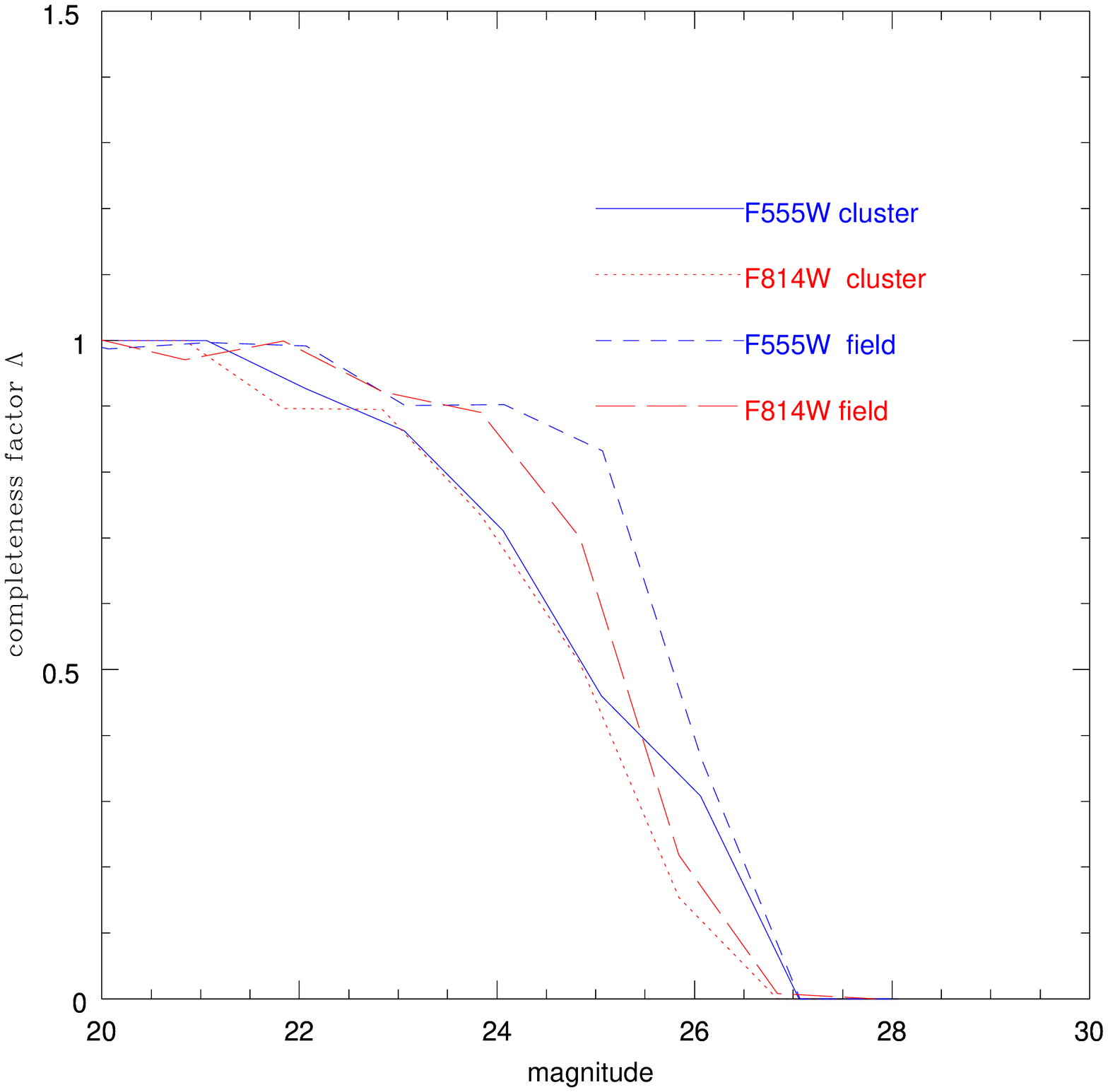}}\\
   }
   \caption{Summary panels for  \object{K~29}. In the top left  panel is the
CM diagram (inner 22'') with a superposed isochrone of log(age)=8.2,
Z=0.003 and E(F555W-F814W)=0.14. The   OGLE bright stars are
indicated by the triangles. The brightest stars settle at F555W=15.
In top right is  the brightness profile (in arbitrary units) fitted
with one of the King (dashed line) and with the Elson functions
(solid line). From this distribution we can determine core and tidal
radius. Vast majority of the stars sets within a radius of 22'' (or
log(r)=1.3). In the bottom left panel the CMD of field stars   is
shown. The field population is taken at about 1.6' from the cluster
center and the area is comparable to  the area of the cluster
(22''). The bottom right panel shows the  completeness factors for
the cluster and field regions as indicated} \label{k29_clu.fig}
\end{figure*}

\begin{figure*}[t]
\parbox{8.7cm}{
\resizebox{8.7cm}{!}{\includegraphics{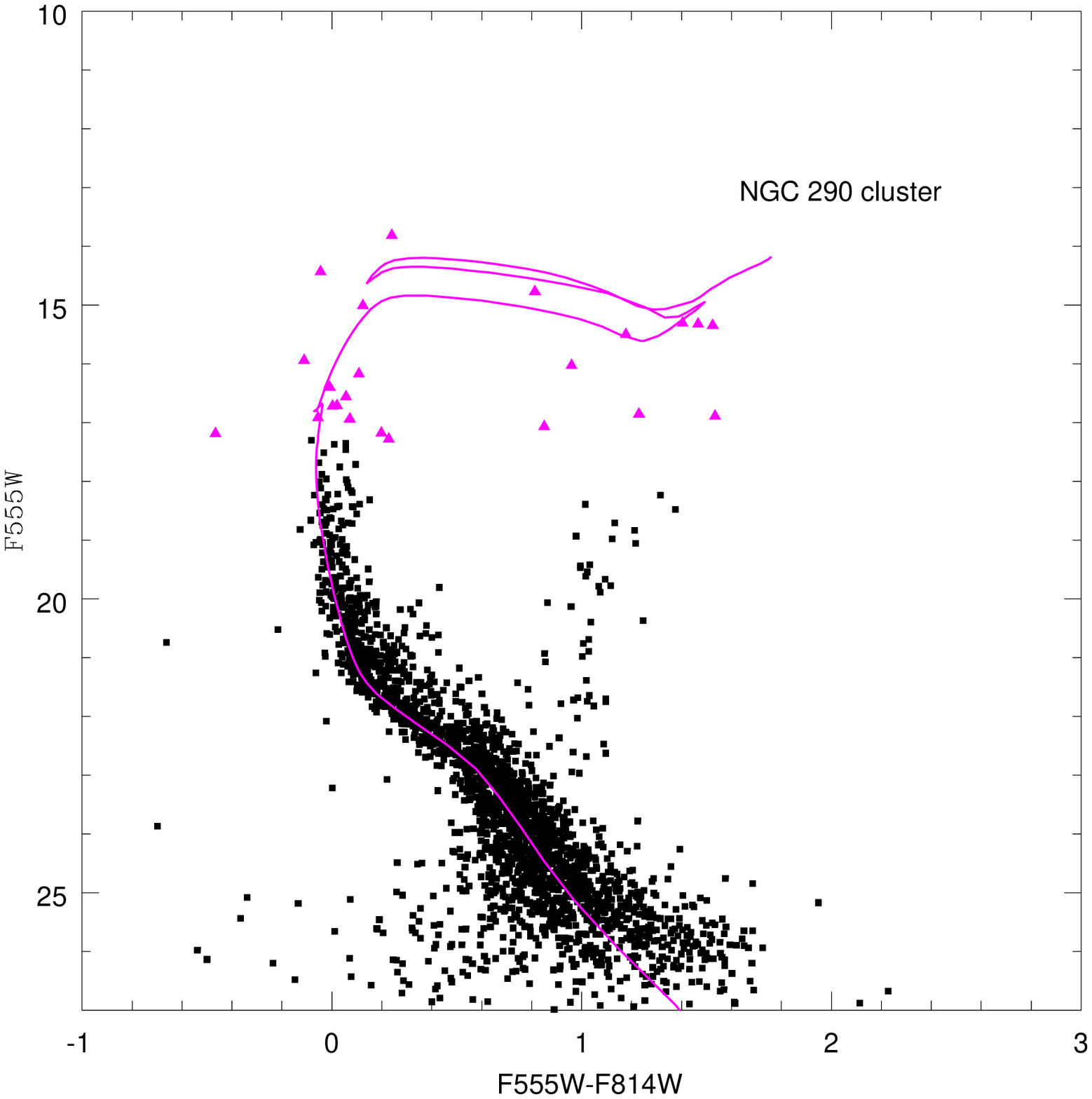}} \\
}
\parbox{8.7cm}{
\resizebox{8.7cm}{!}{\includegraphics{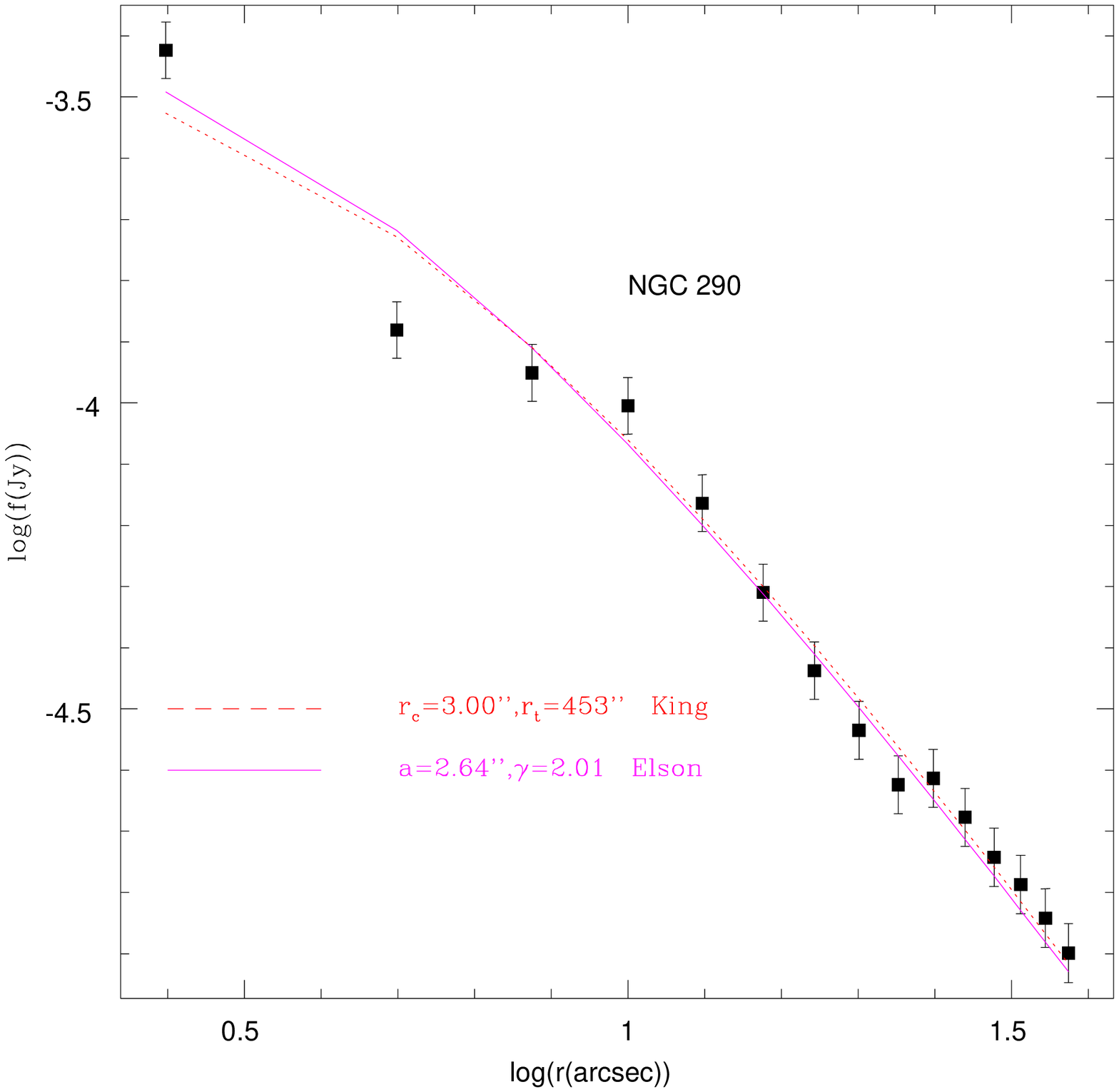}} \\
}
\parbox{8.7cm}{
\resizebox{8.7cm}{!}{\includegraphics{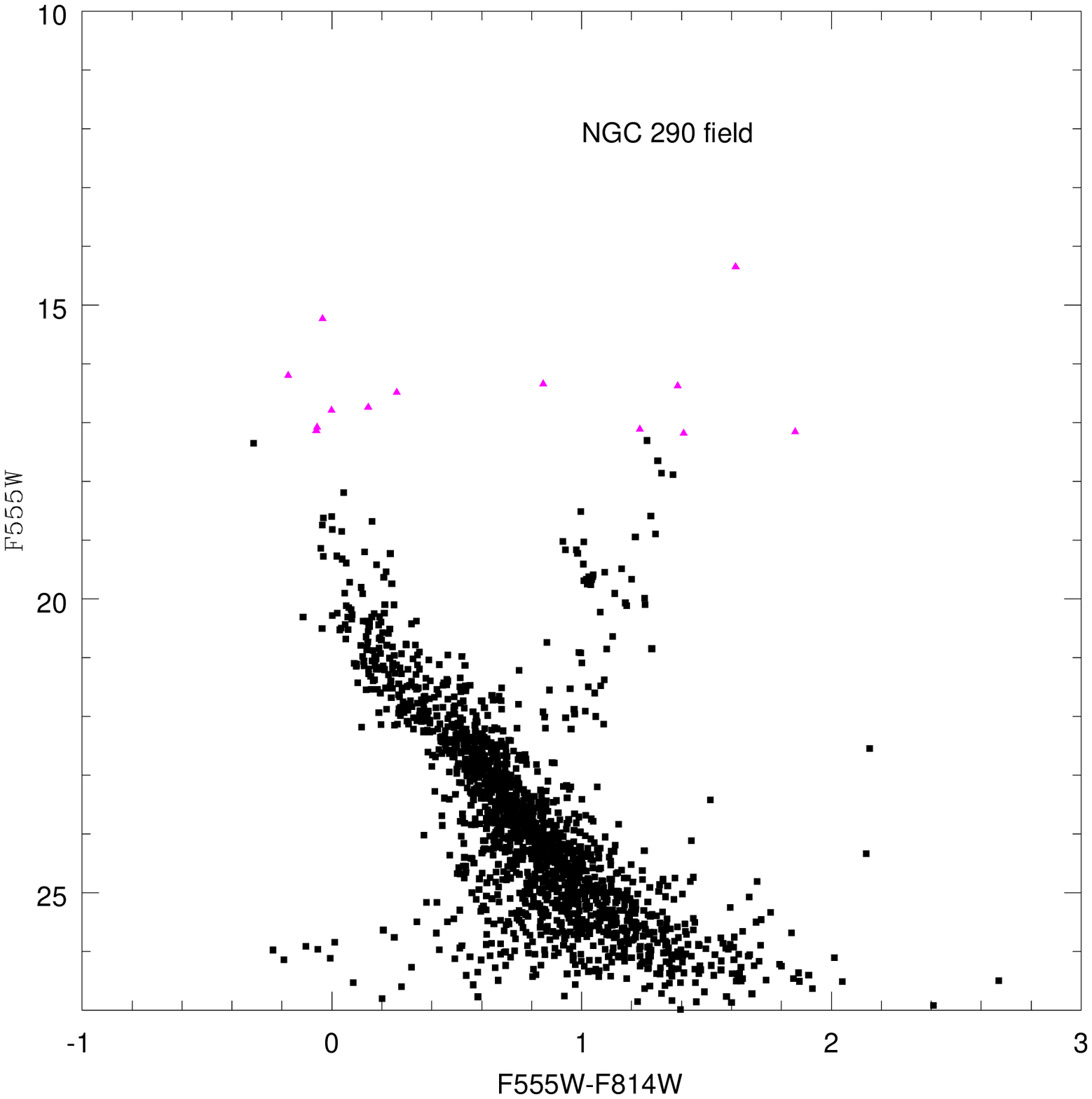}} \\
}
\parbox{8.7cm}{
\resizebox{8.7cm}{!}{\includegraphics{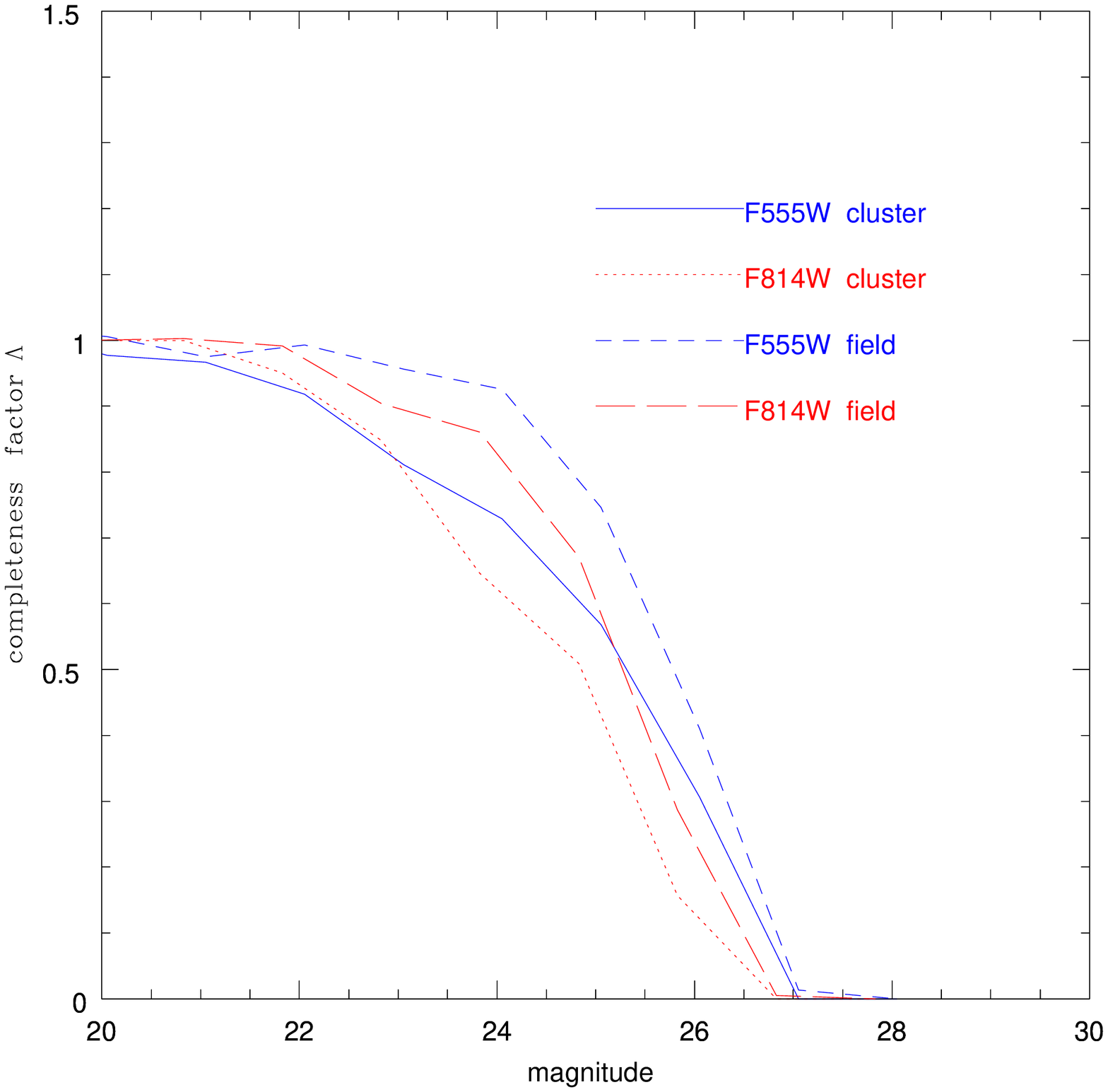}} \\
} \caption{Summary panels for  \object{NGC~290}. In the top left  panel is
the CM diagram (inner radius of 25'') with a superposed isochrone of
log(age)=7.8, Z=0.003,  E(F555W-F814W)=0.15. The   OGLE bright stars
are indicated by the triangles. The brightest stars settle at
F555W=14. In top right is  surface brightness profile (in arbitrary
units) fitted with the King  profile (dashed line) and with the
Elson function (solid line).  From this distribution we can
determine the core and tidal radius. Vast majority of the stars sets
within a radius of 25''. In the bottom left panel the CMD of field
stars  is shown (see text for details). The field population is
taken at about 1.6' from the cluster center and the area is
comparable to  the area of the cluster (25''). The bottom right
panel shows the completeness factors for the  cluster and field
regions as indicated } \label{ngc290_clu.fig}
\end{figure*}


\section{Physical parameters of the clusters}\label{param}

In this section, we first discuss the contamination by the field
population, then we derive the brightness profile, and finally  we
determine  age, metallicity, and IMF slope of each cluster.

\subsection{ Field population contamination}

\begin{figure}
\centering
\resizebox{\hsize}{!}{\includegraphics{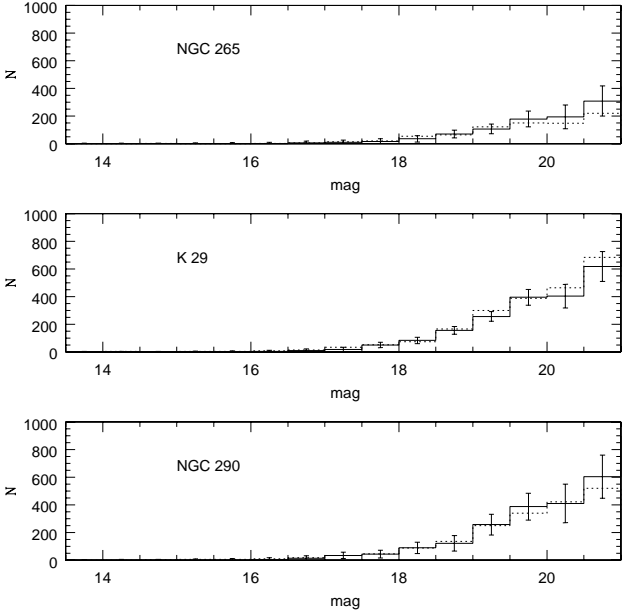}}
\caption{The luminosity functions of HST (solid line) data and OGLE
II (dashed line) data for the field area are compared. OGLE II field
areas are taken at a distance of 1 deg from the clusters. The number
of objects is normalized to an area of $22''$ }
\label{field_comparison}
\end{figure}

The sampling of the field population is a critical issue for surface
profile determination and for the determination of the luminosity
function. The field contamination of the three clusters is
marginally different, (see Fig \ref {field_comparison}) indicating
an inhomogeneous distribution. For this reason and because of the
relatively small size of the field of view of the ACS, we  are
forced to sample the field population  close to the cluster center,
i.e. within about 1.6' from the cluster centers.
%
%
To be sure that the cluster population does not significantly affect
the field sample, we compare the LFs of the field population taken
in this way with the field population taken from OGLE II data at a
distance of about 1 deg. from the cluster centers down to a
magnitude of V=21, where the completeness of the OGLE II data
becomes critical. Fig. \ref {field_comparison} shows that, once that
the LFs are normalized to an equal area, close-by samples of field
population  are statistically comparable to the more distant ones.
This suggests that the contamination by cluster stars inside the
area selected to represent the field  is probably not very relevant.

\subsection{Surface brightness profiles}

In this section, we derive the surface brightness profile of each
cluster. The profiles are then compared with the theoretical ones of
\citet{king1962} and \citet{elson1987} to determine the parameters
of the clusters. First, pixel coordinates of each star are converted
into  the absolute coordinates $\alpha$ and $\delta$ using the task
IRAF STSDAS "xy2rd". Then, the center of a cluster is determined
calculating the luminosity weighted centroid at increasing radii.
The photometric data of the stars that are saturated in the ACS/WFC
observations are taken from OGLE II data, converted to the F555W
pass-band, and finally used to construct the surface brightness
profile.  We are confident that this does not introduce errors in
the derived profiles as  we have already checked the consistency
between  the OGLE II photometry in the V magnitude and the  ACS/WFC
HST V calibration.\\

Many SMC clusters have elliptical isophotes. This is not the case of
the objects studied in this paper, which have only moderate
ellipticity, $1-b/a \sim 0.13--0.3$ \citep{hill2006}. Thanks to
this,  we can use  circular apertures to calculate  the profile.
Because of the high resolution of our images, the surface brightness
given as $log(f)$ inside the i-th circle having radius $b_i$  is
obtained simply by adding up the flux of the stars falling inside
the circle itself:

$$log(f)= 1/{(\pi b_i^2}) \times \sum_j \Lambda_j F_j$$

\noindent where $\Lambda_j$  is the completeness factor inside the
ring and $F_j$ the stellar flux. Crowding can cause a significant
number of stars to be missed by ALLSTAR. Their flux can affect the
surface brightness profile and must be accounted for. This is the
reason why the completeness correction is applied to the flux
determination. We restrict the determination of the surface profile
to the area completely falling inside the small field of view of the
ACS/WFC camera. In all the cases, the profiles have maximum radii of
log(r)$\sim 1.5$ arcsec.


The sky background subtraction is a critical step in the
determination of the surface profile especially in the outer regions
of a cluster.
%
%
The field population is sampled as discussed in the previous
section, the contamination is calculated as the mean flux in a field
area of about 22$''$, and is finally subtracted.


%
Two families of models are fitted on the brightness profiles. The
first one is the tidally truncated  functions of \citet{king1962}:


\begin{equation}
f(x)=\pi r_c^2 k \Bigg[ ln(1+x)-4\frac{(1+x)^{1/2}-1}{(1+x_t)^{1/2}}+\frac{x}{1+x_t}\Bigg]
\end{equation}

\noindent  with $x=(r/r_c)^2$ and $x_t=(r_t/r_c)^2$ where $k$ is a
constant and $r_t$ and $r_c$ are the searched parameters,
specifically $r_t$ is the tidal radius and $r_c$  the core radius.

The second one is the \citet{elson1987} family of functions:

\begin{equation}
f(r)=\frac {2\pi \mu_0 a^2}{\gamma -2}
     \Bigg[1- \Bigg(1+\frac {r^2}{a^2}\Bigg)^{1-\gamma /2}\Bigg]
\end {equation}

\noindent  where $a$ is a scale length and $\gamma$  the power law
index. These models are essentially the same as the \citet{king1962}
models, but they do not assume  tidal truncation. The parameter $a$
is related to the \citet{king1962} core radius $r_c$ by:

$$r_c=a(2^{2/\gamma}-1)^{1/2}$$

\noindent The profile fit is made using the $\chi^2$ minimization.
The uncertainties on the profiles are  calculated as Poisson errors.

\begin{table*}
\caption{Core and tidal radius as determined by the fit with the
\citet{king1962} and \citet{elson1987} functions. }
\begin{center}
\begin{tabular*}{95mm}{l| l c c| c c c c   }
\hline \multicolumn{1}{l|}{Cluster}& \multicolumn{3}{c|}{
\citet{king1962} }&
\multicolumn{4}{c}{ \citet{elson1987}} \\
\hline
          &  $r_c$  & $log(r_t/r_c)$ & $\chi^2$ & $r_c$ & $a$  & $\gamma$  & $\chi^2$  \\
          &(pc)     &                &          & (pc)  & (pc) &           &         \\
\hline
\object{NGC~265}   &   2.61  &  0.75 & 1.06   & 1.98     & 2.85  & 2.63 & 1.06     \\
\object{K~29}      &   2.40  &  2.50 & 1.27   & 2.35     & 2.35  & 2.01 & 1.16  \\
\object{NGC~290}   &   0.87  &  1.50 & 1.55   & 0.76     & 0.76  & 2.01 & 1.53   \\
\hline
\end{tabular*}
\end{center}
\label{radius}
\end{table*}

\subsubsection{Results and comparison with previous work}

Figs. \ref{ngc265_clu.fig}, \ref{k29_clu.fig}, \ref{ngc290_clu.fig}
present the surface brightness profiles of each cluster and
Table~\ref{radius} gives the derived parameters. The core radii
obtained using \citet{elson1987} and \citet{king1962} profiles are
substantially in agreement, as expected because the  main
differences between the models arise at larger radii. The tidal
radius is  quite uncertain, first because it  strongly depends on
the sky subtraction and second, because  our photometry does not
extend to large enough radii. Table~\ref{radius_hill} compares our
results on cluster parameters with those by \citet{hill2006}.
%
%
In comparison with our results, their values of $r_c$ are in
agreement for \object{NGC~265}, but they tend to be  overestimated for the
other two clusters, especially concerning the fit with
\citet{elson1987} profiles. The values of $\gamma$ by
\citet{hill2006} are overestimated for all the clusters.

\begin{figure}
\centering \resizebox{\hsize}{!}{\includegraphics{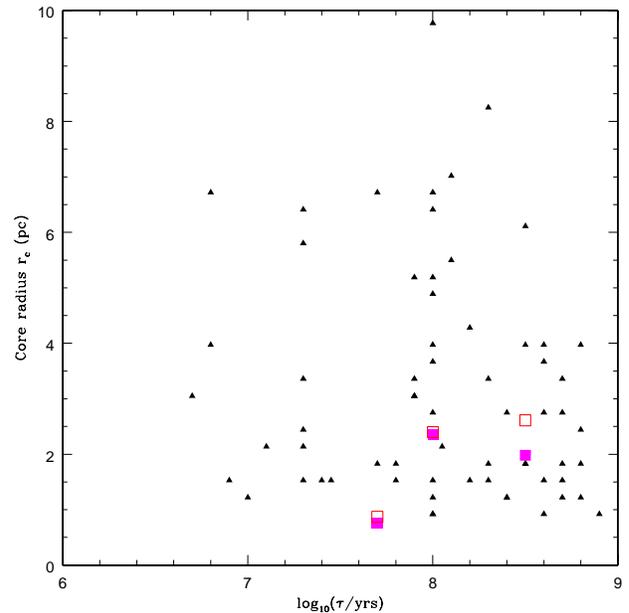}}
\caption{The core radius versus age relationship for SMC clusters
from the sample derived as described in the text. The three clusters studied in this work are
plotted as filled squares for the \citet[][]{elson1987} models and
empty squares for the \citet{king1962} models }
\label{hill_age_rad.fig}
\end{figure}

The question whether MC clusters are tidally limited or not is a
long lasting one. The \citet{king1962} models are expected to
describe bound systems that have isothermal and isotropic stellar
distribution functions and are limited by a strong tidal field. The
\citet{elson1987} models are empirically derived in order  to
reproduce the profiles of young MC clusters. The lack of tidal
cut-off was originally explained with the dynamical youth of those
clusters having extra-tidal stars. In this view, it is possible for a cluster to evolve or not from an  \citet{elson1987} profile to a
\citet{king1962} one, depending on the environment and on the cluster's characteristics. However, on an observational ground, no
clear correlation is found between the age of the clusters and the
presence of tidal cut-off. \citet{hill2006} find that the SMC
clusters are fitted in a satisfactory way by both models, although
the \citet{king1962}  profiles provide a slightly superior fit in
$\sim $ 90\% of the cases. \citet{2005ApJS..161..304M} compare the
\citet{king1962} models, the \citet{elson1987} profiles, and the
modified isothermal spheres based on the \citet{1975AJ.....80..175W}
models. These latter are spatially more extended than the
\citet{king1962} functions, but still include a finite, tidal
cut-off in the density. The authors come to the conclusion that the
un-truncated power-law distributions, \citet{1975AJ.....80..175W}
models, and \citet{king1962} profiles yield  comparable good fits of
MC clusters. \citet{2005ApJS..161..304M} conclude that extended star
envelopes around MC clusters may not be transient features due to
the age, but more probably halos of self-gravitating objects.
For the LMC clusters, \citet{elson1987} find a clear trend that the spread in core radius is an increasing function of the cluster age. The observed trend probably represents a
real structural evolution of the clusters: all clusters form with a relatively small radius and subsequently some of them experience a core expansion, while others do not.  This result is confirmed by other authors \citep
[we quote as the most recent][]{2003MNRAS.338..120M}.
%
%
To
check whether the trend found for the LMC is  a property of SMC clusters as well, we combine the
cluster age determinations derived from isochrone fitting by
\citet{chiosi2006} with the values of the \citet{king1962} core
$r_c$ obtained by \citet{hill2006}. Fig.\ref{hill_age_rad.fig} shows that
a large  spread in core size is present.  From a theoretical point of view, while a core expansion during the evolution is expected as a consequence of the significant and rapid mass loss
by stellar winds, it is less clear why a large scatter is
present, corresponding to a more rapid expansion for some objects. Some mechanisms are proposed in literature which might influence the core evolution, such as  variations in the stellar IMF, the effect of the surrounding tidal field or some external processes (i.e. clusters are not evolving in isolation), the presence of a varying population of primordial binaries. However, at present no satisfactory solution is found.  A wider discussion of this problem is beyond the aims of
this paper. It   can be found in \citet{1991ApJS...76..185E},
\citet{2003MNRAS.338..120M}, \citet{2003MNRAS.343.1025W}.
%
%
It cannot be excluded that  such a large scatter is due to
observational uncertainties on the derived core radius (and ages) of
the clusters. Better quality data are needed before reaching a firm
conclusion.

\begin{table}
\caption{Structural parameters of the clusters as determined by \citet{hill2006}. }
\begin{center}
\begin{tabular}{l| ll |l l  }
\hline \multicolumn{1}{c}{Cluster}& \multicolumn{4}{c}{ Structural
Parameters }
\\
\hline
 &\multicolumn{2}{c|}{\citet{king1962}}& \multicolumn{2}{c}{\citet{elson1987}}          \\
\hline
        & $r_c$ &$r_{90}$&  $r_c$ & $\gamma$ \\
        &(pc)&(pc)&(pc)&\\
\hline
\object{NGC~265} & 2.75 & 11.37& 3.51 & 3.20 \\
\object{K~29}    & 1.22 & 14.58& 5.63 & 6.00 \\
\object{NGC~290} & 1.22 & 14.78& 1.83 & 3.25 \\
\hline
\end{tabular}
\end{center}
\label{radius_hill}
\end{table}

\subsection{Age, metallicity, IMF determinations}

In this section we derive the age, metal content, and IMF  for each
cluster. Table \ref{age.tab} presents the results for all the clusters. Those results are commented in the following Sections.
 Limited to the case of NCG 265, we also compare the CMD and
LF with the predictions from models with and without  overshoot from
the convective core to further contribute to the long lasting debate
on whether convective overshoot is an important phenomenon in
stellar structure and evolution.

\begin{table}
 \caption{The best fit parameters, Age, Z, E${(F555W-F814W)}$  for all the clusters.}
\begin{center}
\begin{tabular}{ l c c c }
\hline
Cluster  &  $log(Age) (yr)$  &  $Z$ &   E${(F555W-F814W)}$  \\
\hline
\multicolumn{4}{c}{ No binaries}\\
\hline
\object{NGC~265}   &    8.5 $\pm$ 0.3  & 0.004$\pm 0.003$     & 0.08 $\pm 0.06$\\
\object{K~29}      &   8.2 $\pm$ 0.2  & 0.003$\pm 0.002$     & 0.14 $\pm 0.08$      \\
\object{NGC~290}   &    7.8 $\pm$ 0.5  & 0.003$\pm 0.002$     & 0.15 $\pm 0.05$\\ 
\hline
\multicolumn{4}{c}{ Binaries}\\
\hline
\object{NGC~265}   &    8.7 $\pm$ 0.2  & 0.004$\pm 0.003$     & 0.08 $\pm 0.06$\\
\object{K~29}      &   8.3 $\pm$ 0.2  & 0.003$\pm 0.002$     & 0.14 $\pm 0.08$ \\
\object{NGC~290}   &    8.0 $\pm$ 0.3  & 0.003$\pm 0.002$     & 0.15 $\pm 0.05$\\

\hline
\end{tabular}
\end{center}
\label{age.tab}
\end{table}

\subsubsection{\object{NGC~265}}


The cluster population is sampled in a circular region of  22''
radius centered on  the cluster. This radius is larger than the core
radius of $\sim 18''$, thus ensuring that the cluster population is
well represented. We sample the field population  at $1.6' $ apart
from the cluster center (see previous sections for a wider
discussion of the field population). To subtract  the field stars
from the cluster LF, we consider a field area of 22$''$. The field
population is statistically subtracted from the LF and the CMD.
 Inside the radius of $22''$  about 1000 cluster stars are
detected down to F555W=25 where the incompleteness reaches the 50\%
level.  Starting from an initial guess obtained from visual
inspection of the CMD, and using the CMD and LF simulator,  by means
of the $\chi^2$ minimization we derive: log(Age)=8.5$\pm 0.3$ yr,
interstellar reddening  E(F555W-F814W)=0.08$\pm0.06$, corresponding
to $A_{F555W}=0.21$ \citep{bedin2005}, and finally  metallicity
Z=$0.004\pm0.003$. All the errors on the determinations of the
parameters are given at the $68\%$ confidence level. The top left
panel of Fig. \ref{ngc265_clu.fig} presents the CMD where the
isochrone for the age and metallicity we have determined is superposed. 
The isochrone gives a good fit of the observational CMD. Different combinations of extinction, metallicity and age (inside the quoted errors) give reasonable fits of the CMD. For instance, when log(Age)=8.4 yr, E(F555W-F814W)=0.1,
and Z=0.007, the main sequence color, the turnoff magnitude, and the luminosity of the loop of the core He-burning stars are well reproduced, while the theoretical loop is too red in comparison with the data. When log(Age)=8.6 yr, E(F555W-F814W)=0.15,
and Z=0.001, then the turnoff and the loop magnitudes and colors are well reproduced, but the colour of the unevolved part of the main sequence,  two magnitudes about below the turnoff is poorly fitted. On the other hand, it is clear that these features of the CMD depend on the details of the stellar models (treatment of convection, mixing length, bolometric corrections, adopted chemical mixture...). For this reason, at some extent, the derived set of
best fitting parameters is  depending on the adopted stellar models.
 The
turn-off mass is about $4 M_{\odot}$. The IMF slope in the  mass
range $M>1M_{\odot}$ turns out to be $\alpha_2=2.5\pm0.5$, while in
the mass range ($0.7 <M<1M_{\odot}$) is $\alpha_1=2.4\pm0.4$,
substantially in agreement with the \citet{kroupa2000} IMF. We
recall that the \citet{kroupa2000} IMF
has $\alpha_1=2.2$ for $0.5M_{\odot}<M<1M_{\odot}$, $\alpha_2=2.7$ for $M>1M_{\odot}$.\\

\begin{figure}
\centering
\resizebox{\hsize}{!}{\includegraphics{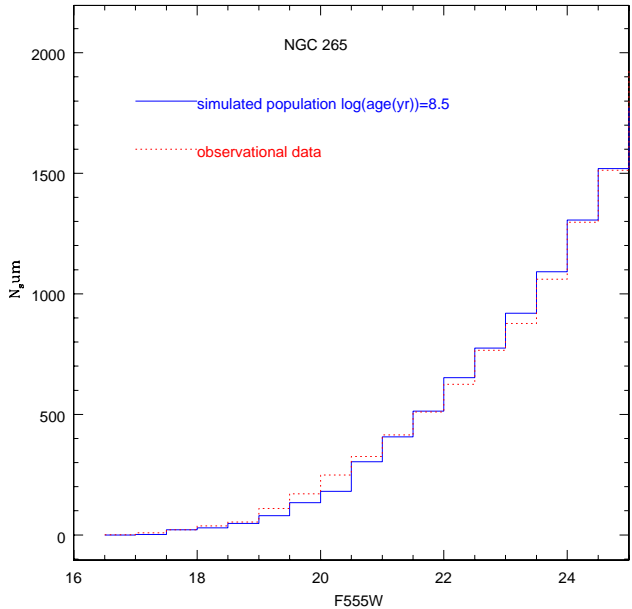}}
\caption{LF of the cluster \object{NGC~265}.  The
best fit is done with a population of log(Age)=8.5 yr}
\label{ngc265f_lum_cfr1.fig}
\end{figure}

When binary stars are included  a slightly older age is found. We
add a percentage of binaries amounting to 30\% of the stars.
Instead of an age of log(Age(yr))=8.5 we find the best fit at an age
of log(Age(yr))=8.7$\pm0.2$.
 As a consequence, the determination of the
exponents of the IMF changes.
%
%
For all the studied clusters, the summary of the IMF determination
is presented in Table \ref{imf.tab}. The quality of the fit is shown
in Fig.\ref{ngc265f_lum_cfr1.fig} where the
observational LF is compared with the best fit simulation.\\

\begin{table}
 \caption{The best fit IMF coefficient in the range
of low masses $\alpha_1$ ($0.7M_{\odot}<M<1M_{\odot}$) and in the range of high
masses $\alpha_2$ ($M>1M_{\odot}$)  is given for the three different clusters.
The 68\% confidence interval is also given ($\sigma_1$ and $\sigma_2$).}
\begin{center}
\begin{tabular}{ l c c c c}
\hline
Cluster  &  $ \alpha_1$  &  $\sigma_1$ &  $ \alpha_2$  &  $\sigma_2$ \\
\hline
\multicolumn{5}{c}{ No binaries}\\
\hline
\object{NGC~265}   &   2.4  &   0.4     &  2.5   & 0.5      \\
\object{K~29}      &   1.8  &   0.2     &  2.7   & 0.3      \\
\object{NGC~290}   &   2.2  &   0.2     &  2.7   & 0.4      \\
\hline
\multicolumn{5}{c}{ Binaries}\\
\hline
\object{NGC~265}   &   2.2  &   0.4     &  2.7   & 0.5      \\
\object{K~29}      &   2.2  &   0.2     &  2.7   & 0.3      \\
\object{NGC~290}   &   2.2  &   0.2     &  2.7   & 0.4      \\

\hline
\end{tabular}
\end{center}
\label{imf.tab}
\end{table}

\subsubsection{\object{K~29}}


We sample the cluster population inside  a  radius of 22'' where
about 700 stars are detected down to $F555W=25$ mag.  As for
\object{NGC~265}, this radius is larger than the core radius.
%
%
The field population of \object{K~29} is located outside the core radius.

\begin{figure}
\centering \resizebox{\hsize}{!}{\includegraphics{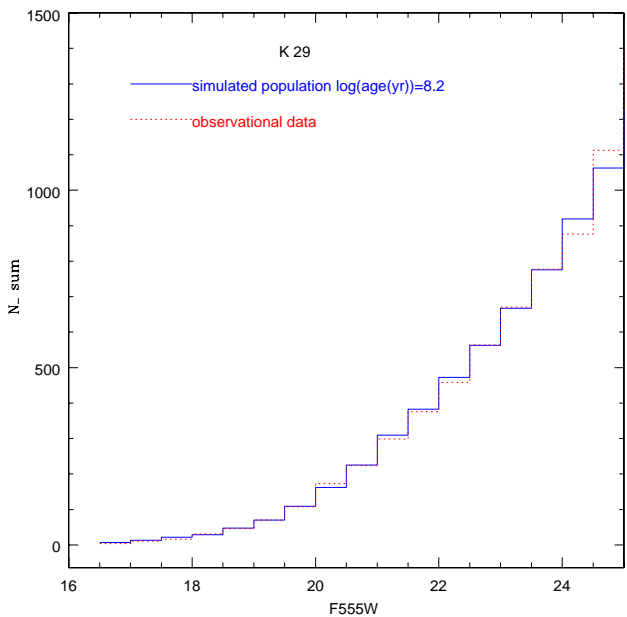}}
\caption{LF  of \object{K~29} fitted with a population of  log(Age)=8.2 yr}
\label{k29pop_isto.fig}
\end{figure}

Fig.\ref{k29_clu.fig} (top left panel) presents the CMD of the
cluster (inside the 22$''$ radius), where bright stars that are
saturated in the ACS/WFC photometry are taken from the OGLE II data,
analogously to the procedure adopted in the determination of the
brightness. The age derived by $\chi^2$ minimization is
log(Age)=$8.2 \pm 0.2$  yr, the metallicity $Z=0.003\pm0.002$, and
the reddening  E(F555W-F814W) =0.14$\pm 0.08$. The turn-off mass is
about $4.5 M_{\odot}$.  The slope of the IMF in the mass interval
from $4 M_{\odot}$ to $1 M_{\odot}$ is $\alpha_2=2.7\pm0.5$, while
in the mass range ($0.7 <M<1M_{\odot}$) is $\alpha_1=1.8\pm0.2$ in
marginal agreement with the Kroupa law.
The LF of the observational data is compared with the best fit model
in Fig.\ref{k29pop_isto.fig}. When binaries are included, the age of
the cluster becomes  log(Age)=$8.3 \pm 0.2$  yr. See the entries of
Table \ref{imf.tab} for the effect of binaries.

\subsubsection{\object{NGC~290}}


Inside a radius of $\sim 25''$ about 660 stars are found. Once more
this radius is larger than the cluster core.  Evolved stars brighter
than  F555W=17.3 mag are beyond the saturation level of the
photometry.  To cope with this we integrate our data for the red
giants with those by OGLE II, whose photometry is converted in the
F555W and F814W pass-bands, as described in the previous sections.
Mounting the two sets of data has been made using equal areas in the
two frames. No stars are missing in the HST photometry at the
turnoff level F555W=18. This ensures that no spurious effects due to
the use of ground-based and HST photometry influences the age
determination.
 The top left panel of Fig.\ref{ngc290_clu.fig} presents the
composite CMD.

\begin{figure}
\centering
\resizebox{\hsize}{!}{\includegraphics{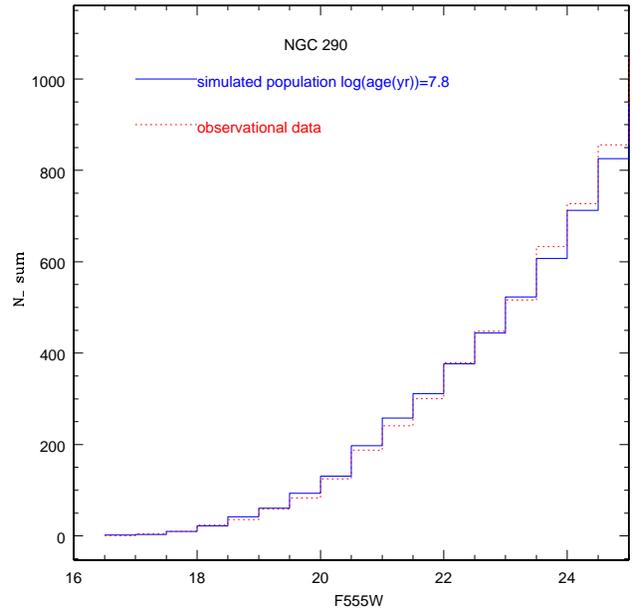}}
\caption{LF of the cluster \object{NGC~290} fitted with a simulated
population having log(Age)=7.8 yr} \label{ngc290pop_isto.fig}
\end{figure}

As for the other two clusters, the preliminary inspection of the CMD
provides estimates of age and metallicity which is then refined by
the Monte Carlo technique. The automatic fit gives log(Age)=7.8$\pm
0.5$ yr with $\chi^2=1.5$,  metallicity  $Z=0.003\pm0.002$, and
E(F555W-F814W) =0.15$\pm 0.05$. The IMF slope gives
$\alpha_1=2.2\pm0.2$ in the mass range $0.7 M_{\odot}\le M \le
1M_{\odot}$, and $\alpha_2=2.7\pm0.4$ in the mass range $M \ge 1
M_{\odot}$. Fig. \ref{ngc290_clu.fig} presents the CMD of \object{NGC~290}
compared with our best fit isochrone. Fig \ref{ngc290pop_isto.fig}
presents the observational LF compared with the best fit model. An
older age is found when 30\% of binary stars are included:
log(Age)=$8.0 \pm 0.3$  yr. Once again see the entries of Table
\ref{imf.tab} for the effect of binaries on the IMF parameters.

\subsection{Revisiting an old question: convective overshoot}

\citet{BM83}, comparing observational and synthetic CMDs of NGC
1866, first noticed that classical stellar models they used
predicted a number of red giants larger than observed, and a smaller
number of main sequence stars in turn as compared to the
observations. The same authors suggested that a more careful
treatment of core convection (i.e. larger convective cores) could
remove the discrepancy. A number of subsequent studies confirmed
this idea and demonstrated that the inclusion of overshooting in the
description of convective motions could reproduce the correct ratio
of the number of evolved stars $N_{\rm R}$ over the number of main sequence stars, $N_{\rm MS}$.
\citet{C89a}, in particular,  clearly
showed that the overshooting scheme, by reducing the ratio between the 
lifetime of the stars in the He-burning and the H-burning phase, $t_{\rm
He}/t_{\rm H}$, offers a good and simple solution to the problem.
This conclusion was also reinforced by \citet{L91}, who analysed the
CMD of NGC 1866 by means of overshooting and classical models.

The classical theory of stellar structure  predicts that in  a star
whenever the condition

\begin{equation}
\nabla_r \leq \nabla_a
\end{equation}

\noindent is violated  the region becomes unstable to convective
motions and fully mixed. This region is surrounded by  stable
radiative layers. This is only an approximation: what happens
physically at the border region passing from unstable to stable
conditions? If we define the Schwarzschild core as the region at
border of which the acceleration of convective bubbles vanishes
there is still a shell above in which the velocity is not zero (the
so-called overshoot region). The thickness, the thermal properties,
and mixing efficiency in the overshooting region are still a matter
of discussion.  In any case, it is worth recalling that first
convective overshoot is  a consequence of the inertia principle, so
that neglecting its existence would not be physically sound, second
it is quite common in Nature \citep{Deardorff69}. As demonstrated in
a number of studies, including numerical simulations
\citep{Freytag96},  the penetration depth of convective elements
into a formally stable region represents a non-negligible fraction
of the size of the unstable zone. In addition to this, there is a
number of astrophysical situations in which the hypothesis of
substantial convective overshoot has been found to offer better and
more elegant solutions than other explanations
\citep[see][]{Be86,C92}. Among others, suffice it to recall here the
so-called mass discrepancy of Cepheid stars \citep{Bw93,C92}. The
first studies of this subject were by
\citet{shaviv1973,prather1974,maeder1975,cloutman1980,bressan1981,stothers1981},
followed over the years by more sophisticated formulations standing
on turbulence theories \citep{Cloutman80,Xiong80} and fluid
hydro-dynamics \citep{Canuto91,Canuto96,UnnoKondo89,Canuto2000}.
The
ballistic description long ago proposed by \citet{bressan1981} is particularly suitable for  the calculation of stellar models. It 
has been proved to best reproduce the numerical results of
laboratory fluid-dynamics simulations \citep{Zahn91}. The
\citet{bressan1981} procedure will be shortly summarized below.

Despite all this, it has been often argued that
 unresolved binaries could mimic the effects of convective overshoot
 and easily
account for the low ratio of evolved stars to main sequence stars
 observed in the young LMC  clusters.
 This hypothesis has been investigated by many authors,
both in open clusters of the Milky Way \citep[see][]{Ca94} and young
clusters of the LMC \citep{C89a,C89b,VaC92} with somewhat
contrasting results. Worth of mention is the  study by \citet{Te99}
of NGC1866, who introducing about $30\,\%$ of binary and using the
classical  models of \citet{Dominguez99} came to the conclusion that
convective overshooting is not required. The subsequent analysis of
\citet{Barmina} of the same cluster carried out on the \citet{Te99}
data, discovered that the correction for completeness applied by
\citet{Te99} was not correct thus invalidating their conclusions. In
addition to this, they performed a systematic analysis of the
effects of binary and stochastic fluctuations in the IMF, and came
to the opposite conclusion: convective overshoot in unavoidable.

Here we have three clusters whose turn-off mass is about 4
$M_{\odot}$ so it might be worth of interest to contribute further
to this subject. In order to check which model better reproduces the data
we need to have  different sets of stellar populations. They are the
two sets with convective overshoot,  i.e. the models by
\citet{girardi2002}  and those by \citet{pietrinferni2004} with a
different prescription so that comparison is possible.

As already recalled, the ballistic description of convective
overshoot by  \citep{bressan1981} is particularly suited to stellar
models. The algorithm adopts a non-local treatment of convection in
the context of the mixing-length theory (MLT)  by \citet{Bohm55}: it
looks for the layers where the velocity of convective elements
(accelerated by the buoyancy force in the formally unstable regions)
gets zero into the surrounding stable regions, then adopts a
suitable temperature stratification in the overshoot regions, and
finally assumes straight mixing over-there. Since the
\citet{bressan1981} formalism makes use of the MLT, it expresses the
mean free path of the convective elements as $l=\Lambda_{\rm c}
\times H_{\rm p}$ where $H_{\rm p}$ is the local pressure scale
height. According to this definition adopted by \citet{bressan1981}
$\Lambda_c$ represents the overshoot distance \emph{across the
Schwarzschild radius} in units of the pressure scale height. In
\citet{bressan1981} $\Lambda_c$ was assumed to be the  same for all
stellar masses. Over the years, this was slightly modified  to treat
overshoot in that particular range of mass which sees the transition
from radiative core H-burning to convective core H-burning and
during this phase the convective core
 starts small, grows to a maximum and then decreases. In this mass
 interval  $\Lambda_c$ is not kept constant but let vary with
mass according to the recipe in \citet{girardi2000}:\\

\noindent
1) $\Lambda_c=0$ for $M \leq 1.0 M_{\odot}$ (when the core is radiative)\\
2) $\Lambda_c=M/M_{\odot}-1.0$ for  $1 M_{\odot}<M \leq 1.5 M_{\odot}$\\
3) $\Lambda_c=0.5$ for  $M > 1.5 M_{\odot}$\\

\noindent The last value stands also for the whole He-burning phase.\\

 \citet{pietrinferni2004} adopt a similar
prescription:\\

\noindent
1) $\Lambda_c=0$ for $M \leq 1.1 M_{\odot}$ (when the core is radiative)\\
2) $\Lambda_c=(M/M_{\odot}-0.9)/4$ for  $1 M_{\odot}<M \leq 1.5 M_{\odot}$\\
3) $\Lambda_c=0.2$ for  $M \geq 1.7 M_{\odot}$\\

\noindent  where we have to keep in mind that $\Lambda_c$ is
"measured" from the Schwarzschild border. Therefore it corresponds
to about half of the \citet{girardi2000} definition. Considered
these premises, the two sets of tracks are practically coincident.
This finding also ensures that using the classical models by
\citet{pietrinferni2004} is safe as the input physics adopted by the
two groups is nearly the same.

\begin{figure}
\centering
\resizebox{\hsize}{!}{\includegraphics{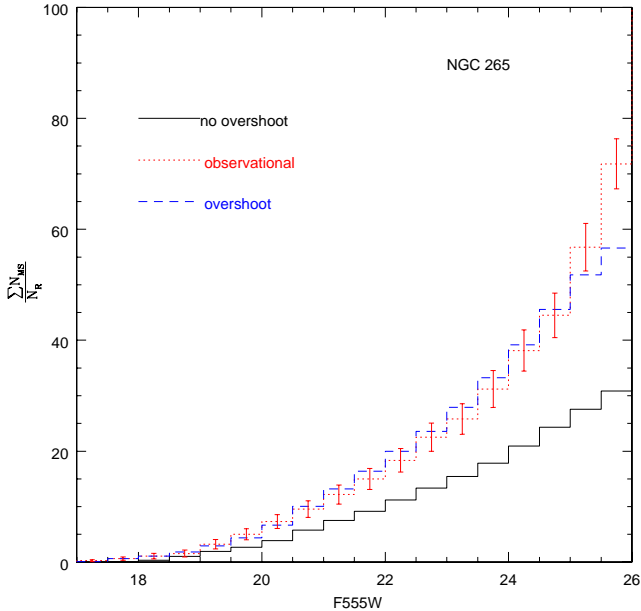}}
\caption{Cluster \object{NGC~265}.  We fit the NILF with canonical models
taken from the BaSTI library and overshooting models from the Padova
Library. Error bars are calculated taking into account photometric
errors on the magnitudes (see text for details) }
\label{isto_over_bert.fig}
\end{figure}


The analysis is  made comparing the integrated luminosity function
of main sequence stars normalized to the number of evolved stars
(thereinafter NILF)  first introduced by \citet{C89a}, which has
been proved to be the only way to effectively discriminate between
the two different evolutionary schemes. It is worth reminding the
reader that the NILF is by definition proportional to the lifetime
ratio $t_{\rm H}/t_{\rm He}$ \citep{C89a}.

The NILF requires that the number of evolved stars truly belonging to
the cluster is known.  The task is not trivial because it requires
an accurate decontamination by field stars. The procedure is  as
follows. First of all, we apply to the CMD of the cluster and field
the correction for completeness and area coverage. Second in the area of the
cluster and field CMD we
perform star counts. We statistically subtract from the cluster
counts a number of evolved stars as large as that in the field. What
remains are the expected cluster evolved stars.
%
%
Finally, we consider only \object{NGC~265} because the small number of giant
stars in \object{NGC~290} and \object{K~29} would introduce too large statistical
uncertainties as it is evident from Table \ref{redgiants} where the
number of giant stars is given. Evolved stars are defined as the
stars brighter than F555W=17 and 16 for \object{K~29} and   \object{NGC~290},
respectively. In the case of \object{NGC~265}, evolved stars are the objects
redder than $(F555W-F814W)\sim 0.4$ and brighter than F555W $= 19$
mag.

\begin{table*}
\caption{Number of evolved stars in the cluster  and field
 CMDs}
\begin{center}
\begin{tabular*}{100mm}{ l c c c c c }
\hline
ID        & Radius($''$)&  Cluster & Field   & Final    & Uncertainty \\
\hline
\object{NGC~265}   & 30          &  51      & 17      &   34     &  $\pm 5$       \\
\object{K~29}      & 22          &  19      & 9       &    10    &  $\pm 3$           \\
\object{NGC~290}   & 25          &  9       & 2       &    7     &  $\pm 3$           \\
\hline
\end{tabular*}
\end{center}
\label{redgiants}
\end{table*}

\begin{figure}
\centering \resizebox{\hsize}{!}{\includegraphics{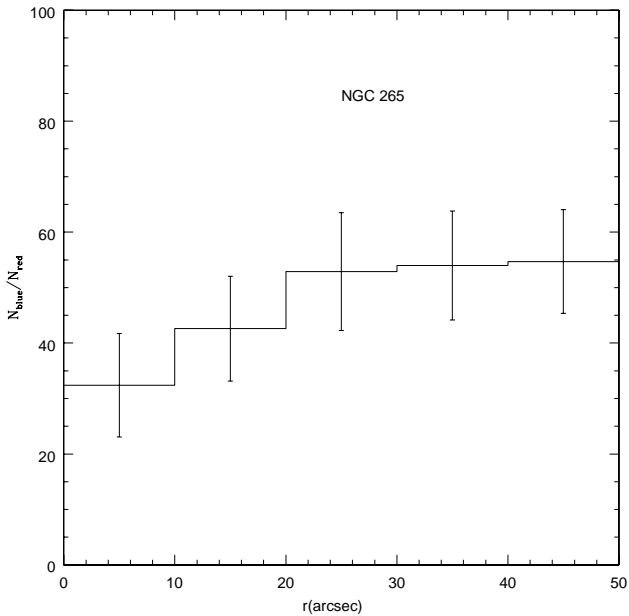}}
\caption{The variation of  the ratio $\sum N_{MS}/N_R$, briefly
indicated as $N_{blue}/N_{red}$,  with the radius
 for \object{NGC~265}  down to the  limit magnitude F555W=25.
The numbers take into account field subtraction and completeness
correction. The error bars are calculated using the Poisson
statistics} \label{redgiants.fig}
\end{figure}

We discuss the uncertainty due to the dependence on the adopted
cluster area since mass segregation cannot be excluded inside the
clusters. Recent estimate of the two-body relaxation time $t_r$
inside MC clusters is given by \citet{2006A&A...452..155K}. They
find values of  $t_r$ of the order
 of $10^7--10^8$ yr inside the core radius, while for the whole clusters
 $t_r$ is of a few Gyr. Although $t_r$ does not exactly correspond to the
time scale for stellar evaporation, this latter is expected to scale
with it. The authors find significant variation of the present-day
mass function exponent with the radius in the inner regions of the
studied clusters as a result of the mass segregation in the core. If
we assume that $t_r$ inside the clusters studied in the present
paper is of the same order and due to  the fact that they have
log(age)$\sim 8$, we come to the conclusion that they can suffer
mass segregation in their core. Fig. \ref{redgiants.fig} shows the
ratio of $\sum N_{MS}/N_R$ with the radius for the studied clusters
for a limiting magnitude F555W=25.
 This ratio  increases with the radius  in the case
of \object{NGC~265}, becoming constant at $30''$.
This is due to the fact that  red giants
are more centrally concentrated than main sequence stars.
%
%
In order to discuss the amount of convective overshoot in the
\object{NGC~265} stars, the region inside a radius of $30''$ should be
considered, to minimize the effects of mass segregation.

The result for the  NILF of \object{NGC~265}, shown in
Fig.\ref{isto_over_bert.fig}, seems to give an observational
function that is intermediate between the two theoretical ones. This
would mean that a certain amount of overshoot is needed to explain
the data.
%
%


To cast light on the reliability of this result, we perform an analysis of the
uncertainties   affecting the NILF.
%
%
We first discuss the effects of the uncertainties due to photometric
errors, mainly  acting on the integrated luminosity function $\sum
N_{MS}$.  From crowding tests we draw the law that relates
photometric errors and magnitudes. Then we compute how many stars
fall in each bin of the histogram with a Monte Carlo simulation,
giving to each star an equal probability to fall in the interval
$v\pm\Delta v$ where $\Delta v$ is the relative photometric error.
We repeat the test 100 times in order to build a sufficient
statistical sample. Then we calculate for each of the bins the mean
value and the root mean square  that we consider as our uncertainty.
The final error is calculated as the partial sum of single bin
errors in the same way as we build the integrated luminosity
function in Fig.\ref{isto_over_bert.fig}.

\begin{figure}
\centering
\resizebox{\hsize}{!}{\includegraphics{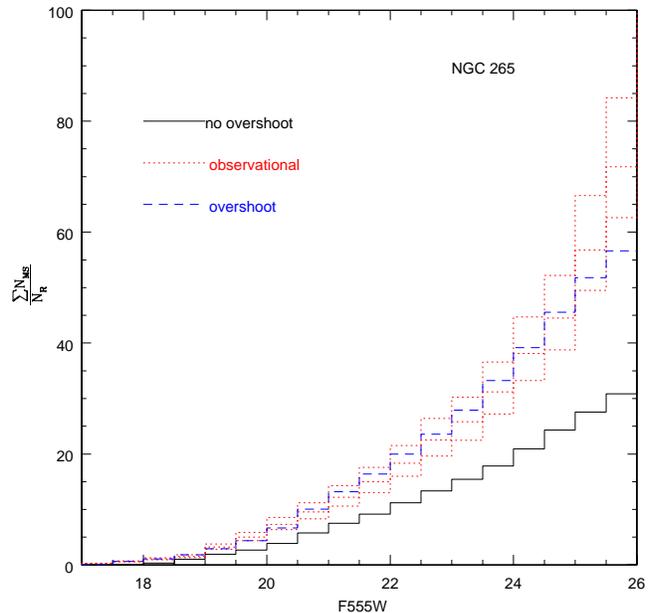}}
\caption{ The variation of  NILF steepness at changing the number of
red giants. The observational NILF (dotted line) is compared with
theoretical counterpart from models with overshoot (dashed line) and
classical models (solid line).  Upper lines are normalized to 30
giants, intermediate to 34, lower lines to 40. We made these
estimates on the consideration that the number of red giants is
known within a certain uncertainty. After the subtraction of field
red giants gives, for NGC 265 we get   $34\pm5$ stars }
\label{isto_over_cfr.fig}
\end{figure}

Second, we evaluate the effect of the statistical uncertainties on
the normalization factor $N_{RG}$.   As expected the steepness of
the NILF is very sensitive to it. The experiment is made for \object{NGC~265}
data looking how the NILF varies at changing the number of red
giants from 30 to 40. The results are shown in
Fig.~\ref{isto_over_cfr.fig}.  We notice first that the photometric
errors do not significantly affect the NILF, second that the observations are in better agreement with overshoot models.
Fig.~\ref{isto_over_cfr.fig}.
%

\section{History of star formation of the surrounding fields}\label{fields}

In this section we derive the star formation rate (SFR) of the field
stars.  Dynamical interaction of the SMC with nearby galaxies, i.e.
the LMC and the Milky Way is a matter of discussion. Precise
determinations of the age at which  intense star formation activity
took place would help to constrain the dynamical problem
\citep{yoshizawa2003}.

In  \citet{chiosi2006} we made use of the OGLE II data to derive the
SFR in the SMC disk. We found evidence of several bursts of star
formation younger  than 2 Gyr together with episodes due to local
phenomena. A burst is found at 200 Myr which can be ascribed  to the
dynamical interaction between the Milky Way and the MCs. The
limiting magnitude of the ground-based photometry did not allow any
firm conclusion concerning older episodes. The deep limiting
magnitude of the present HST-data allows the study of the star
formation rate at  old ages.  In the following, first we present the
method, then we discuss the star formation history of the field
population in the studied regions.

To derive the field SFR, we make use of the whole area of the
companion region of each cluster, 
 and covers  202$\times$101 arcsec.
 The large area of the regions allows a good statistics.
Although in each case we are well outside the core radius of the
associated cluster, some residual cluster contamination might still
be present. This would result in a spurious peak at younger ages.
However, as  we are most interested in the determination of SFR at
old ages, the spurious peak at very young ages would be less of a
problem.

\subsection {The Method}

The detailed description of the adopted method can be found in
\citet{chiosi2006}. Here we recall a few points:

(1) To infer the SFR of a galaxy, theoretical CMDs in different age
ranges are simulated. The simulations include the spread due to
observational photometric errors, reddening, and the effect of
photometric incompleteness. For each age bin, from 10000 to 15000
stars were generated down to the photometric completeness limit. The
synthetic CMDs stand on the sets of stellar tracks, isochrones and
single stellar populations by \citet{girardi2002}. We use N=15
stellar populations, whose ages and metal contents are listed in
Table \ref{pop.tab}.  Populations younger than 100 Myr are
under-sampled and will not be considered in the discussion. The
entries of Table  \ref{pop.tab} also show that the component stellar
populations are chosen in such a way that a suitable chemical
enrichment history is assumed. The one adopted  here is the chemical
history found by \citet{pagel1999}.

(2) The relative contribution of different populations to the total
CMD and luminosity/colour functions to be compared with the
observational ones, in other words the past history of the SFR, is
derived from a minimization algorithm that  stands on the downhill
simplex method \citep{harris2004}. It minimizes the $\chi^2$
function in a parameter-space having N+1 dimensions (the component
stellar populations, here discretized to a finite number N). At each
step of the simplex method, the local $\chi^2$ gradient is derived
and a step in the direction of the gradient is taken, till a minimum
is found. Care is paid to prevent settling of the simplex on local
rather than global minima. Simplex is first started from a random
position, then when a possible solution is obtained, it is
re-started from a position very close to it. Finally, when a minimum
is found,  30000 random directions are searched for a new minimum.

\begin{figure}
\centering \resizebox{\hsize}{!}{\includegraphics{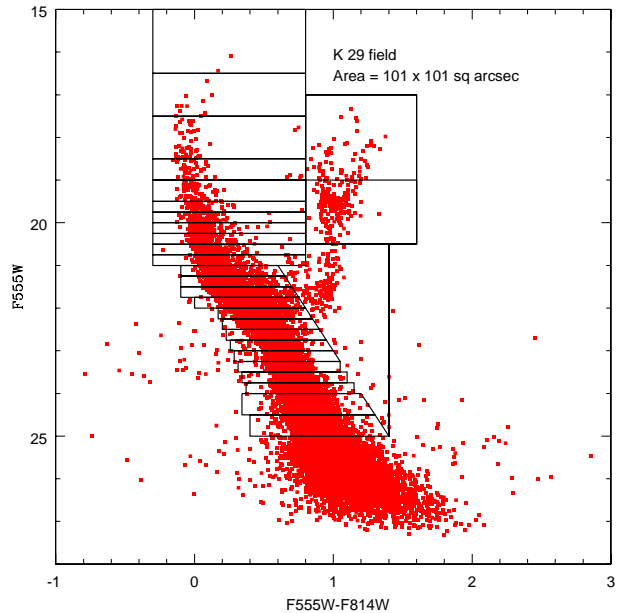}}
\caption{The CM diagram of the field of \object{K~29} with superimposed the
grid adopted to derive the star formation. The same grid is used for
the theoretical simulations. We count all the stars inside each box
both in the synthetic and observational diagrams to construct the
final SF history. The solution of the downhill simplex is the
combination of coefficients (i.e.the mix of synthetic diagrams) that
better fits  the observational diagram} \label{grid.fig}
\end{figure}

(3) To provide the required constraints to the minimization
procedure, we split the  observational CMD into a number of suitable
magnitude-colour bins. Since the fit is mainly based on the main
sequence, 28  bins of variable width are adopted. The bin width is
0.5 mag from 14 to 19 mag, and 0.25 mag from 19 mag to 25 mag. For
the red giant regions only three bins of suitable mag width are used,
i.e.  one for the sub-giant region, and one for the core He-burning
clump, and one for the RGB/AGB stars. The adopted bins are shown in Fig. \ref {grid.fig}. The
simulations are imposed to reproduce the relative percentages of
stars in the various magnitude-colour bins.

\begin{table}
\caption{Ages and metallicities of the synthetic populations in
use.}
\begin{center}
\begin{tabular}{ r c r  c }
\hline
\multicolumn{3}{c}{Age (Gyr)}  & Z  \\
\hline
 0.08 & -- &  0.12   & 0.006 \\
 0.12 & -- &  0.30   & 0.006 \\
 0.30 & -- &  0.40   & 0.006 \\
 0.40 & -- &  0.50   & 0.006 \\
 0.50 & -- &  0.60   & 0.006 \\
 0.60 & -- &  0.80   & 0.006 \\
 0.80 & -- &  1.00   & 0.006 \\
 1.00 & -- &  2.00   & 0.001  -- 0.006 \\
 2.00 & -- &  3.00   & 0.001  -- 0.003 \\
 3.00 & -- &  4.00   & 0.001  -- 0.003 \\
 4.00 & -- &  5.00   & 0.001  -- 0.003 \\
 5.00 & -- &  6.00   & 0.001  -- 0.003 \\
 6.00 & -- &  8.00   & 0.001  -- 0.001 \\
 8.00 & -- & 10.00   & 0.001  -- 0.001 \\
10.00 & -- & 12.00   & 0.001  -- 0.001 \\
\hline
\end{tabular}
\end{center}
\label{pop.tab}
\end{table}

\subsection{Results and discussion}\label{discussion}

\begin{figure}
\centering
\resizebox{\hsize}{!}{\includegraphics{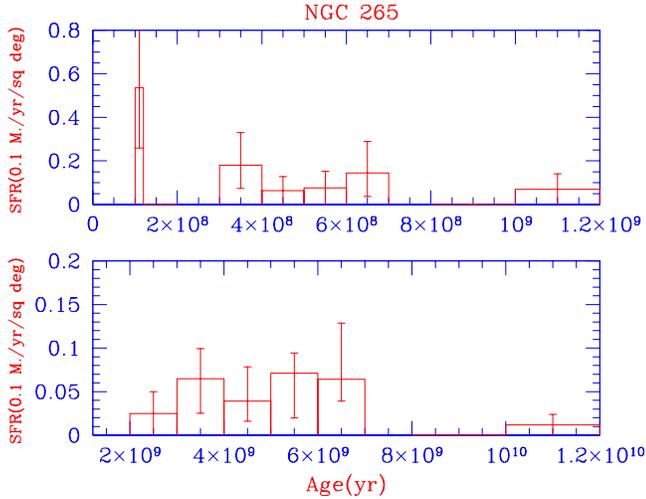}}
\caption{Star formation history for the field of  \object{NGC~265}. The age of
the associated cluster corresponds to the burst of star formation at
300 Myr.  The SFR  is normalized to the area}
\label{out_field_mass.fig}
\end{figure}

\begin{figure}
\centering \resizebox{\hsize}{!}{\includegraphics{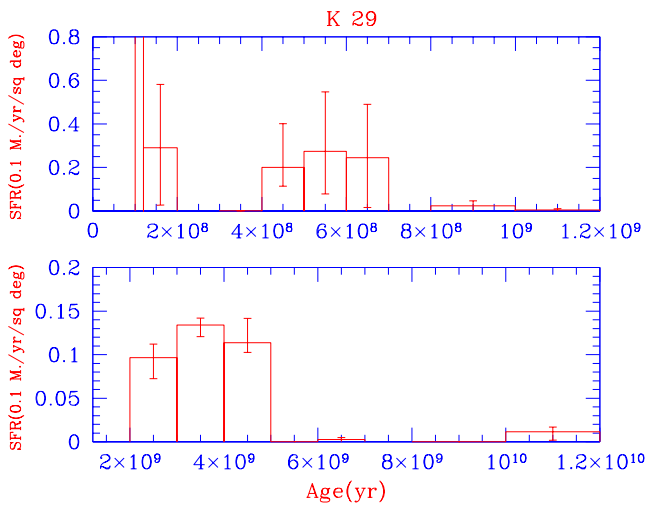}}
\caption{Star formation history for the field of \object{K~29}. We can see
bursts of star formation at ages 200-400 Myr, and between 3 Gyr and 5 Gyr. The
age of the associated cluster corresponds to the first peak. The SFR
is normalized to the area} \label{out_field_k29.fig}
\end{figure}

\begin{figure}
\centering \resizebox{\hsize}{!}{\includegraphics{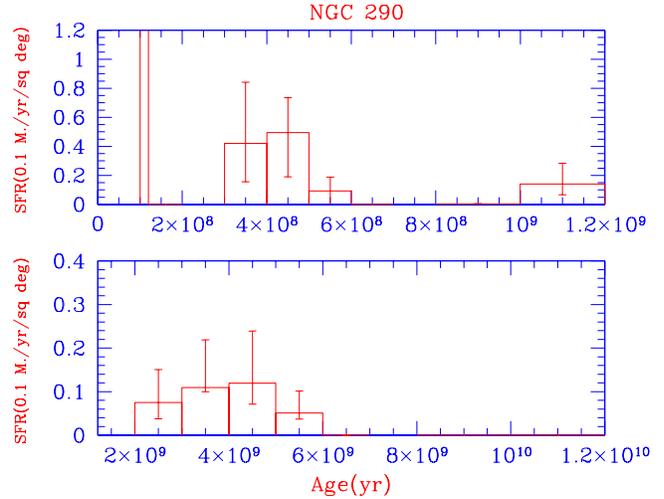}}
\caption{Star formation history for the field of \object{NGC~290}. Main star
formation episodes are at 300-400 Myr, and between 3 Gyr and 6 Gyr. Those
episodes are common to the three areas suggesting a global
triggering mechanism. The rate is normalized to the area}
\label{ngc290_out_field.fig}
\end{figure}

Star formation in the three fields shows several  common features
(see Figs. \ref{out_field_mass.fig}, \ref{out_field_k29.fig}, and
\ref{ngc290_out_field.fig}). In all the fields, the star formation
was not continuous but proceeded in a number of bursts taking place
at 0.3-0.4 Gyr, and between 3 Gyr and  6 Gyr ago.

The bursts at 3 and 0.4 Gyr are temporally coincident with past
peri-galactic passages of the SMC about the Milky Way. It is well
known that the young field population presents an asymmetric
structure biased toward the eastern  LMC-facing side of the SMC.
\citet{crowl2001} find the same trend among the  SMC populous
clusters: those toward the eastern side tend to be younger and  more
metal rich than  those on the western side. This is interpreted as
the effect of the perturbations developed by  the interaction of
LMC-SMC-MW \citep[see][for a wider discussion]{chiosi2006}. This is
the kind of reshaping that a low mass disc galaxy ought to undergo
after a few passages around a more massive galaxy
\citep{pasetto2003}.


While a larger consensus is reached in literature concerning the SFR
at young ages and the influence of the interaction between the MCs
and the Milky Way, still controversial results are obtained
concerning the old population. In a pioneering work based on
photographic plates, \citet{gardiner1992} found that the bulk of the
stellar population in the SMC is about 10 Gyr old.
%
%
\citet{harris2004} find that  50\% of the stars  ever born in the
SMC formed prior to 8.4 Gyr ago. More precisely, the SMC formed
relatively few stars between 8.4 and 3 Gyr ago, but there was a rise
in the mean SFR during the most recent 3 Gyr with bursts at ages of
2.5, 0.4, and 0.06 Gyr.
%
%
\citet{dolphin2001} find that the star formation has proceeded in a
more
 continuous way, with  a main  episode between 5 and 8 Gyr.
\citet{Noel2006} find a strong gradient with the galacto-centric
distance and  no significant star formation activity in the SMC  at
ages older than 6-9 Gyr.

\begin{figure}
\centering
\resizebox{\hsize}{!}{\includegraphics{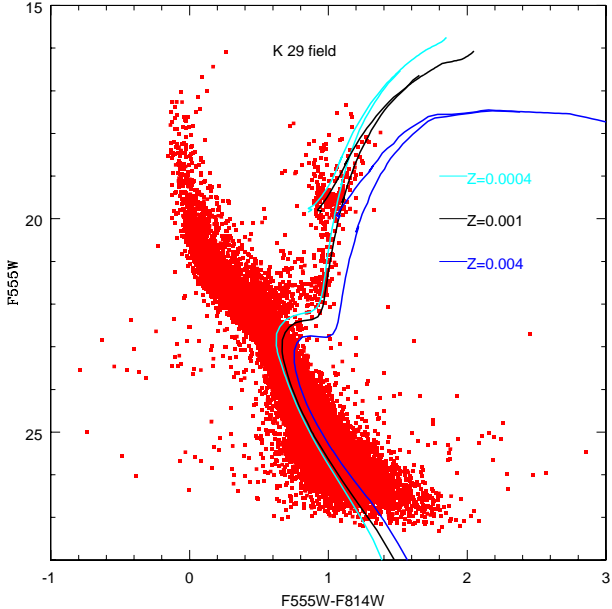}}
\caption{The CM diagram of the field of \object{K~29} with superimposed three
isochrones of 10 Gyr at different metallicity. The sub-giant branch
width is not compatible with the presence of a significant
population as old as or older than 10 Gyr when the metallicity is
ranging from Z=0.0004 to 0.004} \label{age_old}
\end{figure}

The present result  suggests a low efficiency of the star formation
at old epochs. Using Monte Carlo simulations, we derive the
uncertainty on this result. Small changes in the assumed
line-of-sight SMC depth, extinction, age-metallicity relation result
in an uncertainty of at least $\pm 2 $ Gyr on the age of the
components older than 3 Gyr. 
This is consistent with the magnitude bin size adopted to study the CMD. In fact  a 0.25 mag uncertainty on
the turnoff magnitude results in 2 Gyr age difference  for a
component of 5 Gyr, and Z=0.004. If we adopt a different metallicity, in  the range
0.004 to 0.001, the age-metallicity degeneracy is responsible of an
additional uncertainty of about 0.5 Gyr.


To assess whether  a significant component as old as or even older
than 10 Gyr is present in our fields, we compare the isochrones at
changing metallicity with the CMD. The result is plotted in Fig.
\ref{age_old}. The sub-giant branch width is not compatible with the
presence of a significant component as old as  or older than 10 Gyr.

 This is in agreement with the age distribution of the clusters.
In fact the SMC is known  to have  at least six populous clusters of
intermediate age,  namely in the range 5-9 Gyr, but only one true
old object  (NGC 121) having an age $>$ 10 Gyr  is known in this
galaxy \citep{stryker1985, dolphin2001}. The SFR of the field
population in the LMC at older ages is not inconsistent with these
results. In fact the majority of the authors find a strong
population at intermediate ages (between 2-6 Gyr) \citep [we quote
among others][]{elson1997, geha1998, harris1999, harris2004,
olsen1999, dolphin2000a, javiel2005}, but only in a few cases
enhancements at older ages are found. \citet{vallenari1996} find
that in  the regions East and South-East of the Bar the star
formation had a sort of  enhancement   about 6-8 Gyr ago.
\citet{harris2004} derive two main episodes of star formation  at
epochs younger than 2 Gyr and older of 7 Gyr respectively, with a
quiescent period in between. \citet{smecker-hane2002} find that the
Disk SFR has been relatively smooth and continuous over the last 13
Gyr, while the Bar was dominated by SF episodes at intermediate
ages, more precisely from 4 to 6 Gyr and 1 to 2 Gyr ago.

How can this be related to the interaction between the MCs and our
Galaxy? Over the years our understanding of the problem   has been
much improving thanks to studies aimed at reproducing the properties
of LMC and SMC by shaping their SFH by means of the mutual dynamical
interactions in the triplet LMC+SMC+MW. However, it is still not
clear whether the MCs have existed in a binary state for most of
their history \citep{1984IAUS..108..115F, gardiner1994} or if their
strong interaction happened  just recently
\citep{mathewson1986,1994AJ....107.2055B, bekki2005, yoshizawa2003}.
The solution is not unique owing to uncertainty in the initial
conditions, and furthermore in many cases the MCs cannot keep their
binary status  for more than about 5-6 Gyr \citep{bekki2005,
yoshizawa2003}. \citet{bekki2004} find that it is  unlikely that the LMC and the SMC were a binary system more than 6 Gyr ago. \citet{bekki2005}  pointed out that gravitational interactions do not necessarily influence in the same way the LMC and the SMC  and  that cluster and field star  formed in different epochs.
Since the LMC formed far from the Milky Way (MW), no tidal trigger of the SF at old times is expected in this galaxy, while because of its smaller mass and its proximity, the  SMC SFR was more influenced by MW tidal trigger. No cluster formation in the  {LMC} took place as a consequence  of the first perigalactic passage, 6.8 Gyr ago, when instead the field star formation was enhanced.
The present  results concerning the SMC, suggest that at old ages, either there was no gravitational interaction between the two galaxies in agreement with \citet{bekki2004}, or the
tidal interaction between the MCs,  was not strong enough to
trigger star formation episodes in the SMC.


\section{Summary and conclusions}
\label{conclusion}
In this paper we have studied three clusters and companion fields of
the SMC observed by E. Olszewski in year 2004 as part of the ACS/WFC
program. The three clusters are located at the east side of HI
super-shell 37A  whose borders overlap  a region  of active star
formation, most likely caused  by the propagation of the gas
pressure wave \citep{staveley1997,stanimirovic1999}. The ages
assigned to the clusters on the basis of their CMD and luminosity
function seem to be too old to be related to the dynamical age
of the shell \citep{stanimirovic1999}. \\

The ACS/WFC frame of each cluster is split  in two parts, one
containing the cluster itself and the other containing the companion
field. Images are reduced with the usual DAOPHOT/ALLSTAR packages,
completeness is calculated separately for the cluster  and  the
field. We find the completeness limits
(50\% complete) for magnitudes F555W=25 for the clusters and F555W=26 for the fields.\\

In the ACS/WFC field containing the  cluster we select a circular
region inside which most of the cluster stars are located and derive
the CMD and integrated luminosity function. The luminosity function
and CMD  are then compared to  a library of synthetic luminosity
functions and isochrones  covering a large range of ages and
metallicities. This yields   a provisional value of the age and
metal content best reproducing these two constraints. The results
are then refined with the aid of population synthesis technique and
the $\chi^2$-minimization algorithm. We find the following ages and
metallicities: for the cluster \object{NGC~265} the age is
log(Age)=$8.5\pm0.3$ yr and the metallicity is Z=$0.004\pm0.003$(corresponding to $[Fe/H]\sim -0.62$); for
the cluster \object{K~29} the age is log(Age)=$8.2\pm0.2$ yr and the
metallicity is  Z=$0.003\pm0.002$(corresponding to $[Fe/H]\sim -0.75$); for the
cluster \object{NGC~290} the  age is log(Age)=$7.8\pm0.5$ yr and the
metallicity is Z=$0.003\pm0.002$(corresponding to $[Fe/H]\sim -0.75$).\\

We also estimate the slope  of the IMF, especially at low masses,
between $0.7M_{\odot}$ and $1M_{\odot}$. We check whether the Kroupa
exponent, $\alpha_1=2.2$, is recovered in this range. We see that all
the clusters have the minimum  of $\chi^2$ function for $\alpha_1$
close but not identical to the Kroupa value.  The values are
$\alpha_1=2.4\pm0.4$ for \object{NGC~265}, $\alpha_1=1.8\pm0.2$ for \object{K~29},
$\alpha_1=2.2\pm0.2$ for \object{NGC~290}. The differences are not significant at
the 68\% confidence level.\\

Limited to  \object{NGC~265}, we also tested the theory of convective
overshoot: we fitted the experimental NILF with two theoretical
models, one including overshoot and the other without it taken from the BaSTI library.  We can say that
a certain amount of overshoot is needed in order to fit the experimental data.\\

Then we determine the past star formation history  for the companion
fields. The three determinations of the SF history for the three
fields  show several common features suggesting a common triggering
mechanism over  the scale length of interest here ($\sim 700$ pc).
Main episodes are at 300-400 Myr, and between 3 Gyr and 6 Gyr. However the SFR
was not very efficient at older ages. This result is in agreement
with the cluster age distribution in the SMC and with the SFR at old
ages in the LMC. This suggests that at old ages, the tidal
interaction between the MCs, if any, was not strong enough to
trigger star formation episodes in both galaxies Finally,  we point
out  that our determination of the past  SF history is not precise
for  ages younger than about 100 Myr.

\begin{acknowledgements}
The authors thank  C. Chiosi  for reading   the preliminary version
of the manuscript, and G. Bertelli for many useful discussions and
clarifications on the synthetic CMD generator. This study has been
financially supported by the Department of Astronomy of the Padova
University and the Padova Astronomical Observatory of the National
Institute for Astrophysics (INAF).
\end{acknowledgements}

\bibliographystyle{aa}

\bibliography{vallenari_6834}

\begin{thebibliography}{95}
\expandafter\ifx\csname natexlab\endcsname\relax\def\natexlab#1{#1}\fi

\bibitem[{{Barmina} {et~al.}(2002){Barmina}, {Girardi}, \& {Chiosi}}]{Barmina}
{Barmina}, R., {Girardi}, L., \& {Chiosi}, C. 2002, \aap, 385, 847

\bibitem[{{Becker} \& {Mathews}(1983)}]{BM83}
{Becker}, S. \& {Mathews}, J. 1983, AJ, {\bf 270}, 155

\bibitem[{{Bedin} {et~al.}(2005){Bedin}, {Cassisi}, {Castelli}, {Piotto},
  {Anderson}, {Salaris}, {Momany}, \& {Pietrinferni}}]{bedin2005}
{Bedin}, L.~R., {Cassisi}, S., {Castelli}, F., {et~al.} 2005, \mnras, 357, 1038

\bibitem[{{Bekki} \& {Chiba}(2005)}]{bekki2005}
{Bekki}, K. \& {Chiba}, M. 2005, \mnras, 356, 680

\bibitem[{{Bekki} {et~al.}(2004){Bekki}, {Couch}, {Beasley}, {Forbes}, {Chiba},
  \& {Da Costa}}]{bekki2004}
{Bekki}, K., {Couch}, W.~J., {Beasley}, M.~A., {et~al.} 2004, \apjl, 610, L93

\bibitem[{{Bertelli} {et~al.}(1986{\natexlab{a}}){Bertelli}, {Bressan},
  {Chiosi}, \& {Angerer}}]{Be86}
{Bertelli}, G., {Bressan}, A., {Chiosi}, C., \& {Angerer}, K.
  1986{\natexlab{a}}, A\&AS, {\bf 66}, 191

\bibitem[{{Bertelli} {et~al.}(1986{\natexlab{b}}){Bertelli}, {Bressan},
  {Chiosi}, {Mateo}, \& {Wood}}]{Bw93}
{Bertelli}, G., {Bressan}, A., {Chiosi}, C., {Mateo}, M., \& {Wood}, P.
  1986{\natexlab{b}}, ApJ, {\bf 412}, 160

\bibitem[{{B\"ohm-Vitense}(1958)}]{Bohm55}
{B\"ohm-Vitense}, E. 1958, Z. Astroph., {\bf 46}, 108

\bibitem[{{Bressan} {et~al.}(1981){Bressan}, {Chiosi}, \&
  {Bertelli}}]{bressan1981}
{Bressan}, A.~G., {Chiosi}, C., \& {Bertelli}, G. 1981, \aap, 102, 25

\bibitem[{{Byrd} {et~al.}(1994){Byrd}, {Valtonen}, {McCall}, \&
  {Innanen}}]{1994AJ....107.2055B}
{Byrd}, G., {Valtonen}, M., {McCall}, M., \& {Innanen}, K. 1994, \aj, 107, 2055

\bibitem[{{Canuto}(2000)}]{Canuto2000}
{Canuto}, V.~M. 2000, \apjl, 534, L113

\bibitem[{{Canuto} {et~al.}(1996){Canuto}, {Goldman}, \&
  {Mazzitelli}}]{Canuto96}
{Canuto}, V.~M., {Goldman}, I., \& {Mazzitelli}, I. 1996, ApJ, {\bf 473}, 550

\bibitem[{{Canuto} \& {Mazzitelli}(1991)}]{Canuto91}
{Canuto}, V.~M. \& {Mazzitelli}, I. 1991, ApJ, {\bf 370}, 295

\bibitem[{{Caputo} {et~al.}(1999){Caputo}, {Marconi}, \& {Ripepi}}]{caputo1999}
{Caputo}, F., {Marconi}, M., \& {Ripepi}, V. 1999, \apj, 525, 784

\bibitem[{{Carraro} {et~al.}(1994){Carraro}, {Chiosi}, {Bressan}, \&
  {Bertelli}}]{Ca94}
{Carraro}, G., {Chiosi}, C., {Bressan}, A., \& {Bertelli}, G. 1994, A\&AS, {\bf
  103}, 375

\bibitem[{{Chiosi} {et~al.}(1992){Chiosi}, {Bertelli}, \& {Bressan}}]{C92}
{Chiosi}, C., {Bertelli}, G., \& {Bressan}, A. 1992, ARA\&A, {\bf 30}, 235

\bibitem[{{Chiosi} {et~al.}(1989{\natexlab{a}}){Chiosi}, {Bertelli}, {Meylan},
  \& {Ortolani}}]{C89a}
{Chiosi}, C., {Bertelli}, G., {Meylan}, G., \& {Ortolani}, S.
  1989{\natexlab{a}}, A\&A, {\bf 219}, 167

\bibitem[{{Chiosi} {et~al.}(1989{\natexlab{b}}){Chiosi}, {Bertelli}, {Meylan},
  \& {Ortolani}}]{C89b}
{Chiosi}, C., {Bertelli}, G., {Meylan}, G., \& {Ortolani}, S.
  1989{\natexlab{b}}, A\&AS, {\bf 78}, 89

\bibitem[{{Chiosi} {et~al.}(2006){Chiosi}, {Vallenari}, {Held}, {Rizzi}, \&
  {Moretti}}]{chiosi2006}
{Chiosi}, E., {Vallenari}, A., {Held}, E.~V., {Rizzi}, L., \& {Moretti}, A.
  2006, \aap, 452, 179

\bibitem[{{Cloutman} \& {Whitaker}(1980{\natexlab{a}})}]{Cloutman80}
{Cloutman}, L. \& {Whitaker}, R.~W. 1980{\natexlab{a}}, ApJ, {\bf 237}, 900

\bibitem[{{Cloutman} \& {Whitaker}(1980{\natexlab{b}})}]{cloutman1980}
{Cloutman}, L.~D. \& {Whitaker}, R.~W. 1980{\natexlab{b}}, \apj, 237, 900

\bibitem[{{Crowl} {et~al.}(2001){Crowl}, {Sarajedini}, {Piatti}, {Geisler},
  {Bica}, {Clari{\' a}}, \& {Santos}}]{crowl2001}
{Crowl}, H.~H., {Sarajedini}, A., {Piatti}, A.~E., {et~al.} 2001, \aj, 122, 220

\bibitem[{{Da Costa} \& {Hatzidimitriou}(1998)}]{dacosta1998}
{Da Costa}, G.~S. \& {Hatzidimitriou}, D. 1998, \aj, 115, 1934

\bibitem[{{Deardorff} {et~al.}(1969){Deardorff}, {Willis}, \&
  {Lilly}}]{Deardorff69}
{Deardorff}, J., {Willis}, G., \& {Lilly}, D. 1969, Fluid Mech., {\bf 35}, 7

\bibitem[{{Dolphin}(2000{\natexlab{a}})}]{dolphin2000a}
{Dolphin}, A.~E. 2000{\natexlab{a}}, \pasp, 112, 1397

\bibitem[{{Dolphin}(2000{\natexlab{b}})}]{dolphin2000b}
{Dolphin}, A.~E. 2000{\natexlab{b}}, \mnras, 313, 281

\bibitem[{{Dolphin} {et~al.}(2001){Dolphin}, {Walker}, {Hodge}, {Mateo},
  {Olszewski}, {Schommer}, \& {Suntzeff}}]{dolphin2001}
{Dolphin}, A.~E., {Walker}, A.~R., {Hodge}, P.~W., {et~al.} 2001, \apj, 562,
  303

\bibitem[{{Dominguez} {et~al.}(1999){Dominguez}, {Chieffi}, {Limongi}, \&
  {Straniero}}]{Dominguez99}
{Dominguez}, I., {Chieffi}, A., {Limongi}, M., \& {Straniero}, O. 1999, ApJ,
  {\bf 524}, 226

\bibitem[{{Elson} {et~al.}(1998){Elson}, {Sigurdsson}, {Davies}, {Hurley}, \&
  {Gilmore}}]{El98}
{Elson}, A., {Sigurdsson}, S., {Davies}, M., {Hurley}, J., \& {Gilmore}, G.
  1998, MNRAS, {\bf 300}, 857

\bibitem[{{Elson}(1991)}]{1991ApJS...76..185E}
{Elson}, R.~A.~W. 1991, \apjs, 76, 185

\bibitem[{{Elson} {et~al.}(1987){Elson}, {Fall}, \& {Freeman}}]{elson1987}
{Elson}, R.~A.~W., {Fall}, S.~M., \& {Freeman}, K.~C. 1987, \apj, 323, 54

\bibitem[{{Elson} {et~al.}(1997){Elson}, {Gilmore}, \& {Santiago}}]{elson1997}
{Elson}, R.~A.~W., {Gilmore}, G.~F., \& {Santiago}, B.~X. 1997, \mnras, 289,
  157

\bibitem[{{Freytag} {et~al.}(1996){Freytag}, {Ludwig}, \&
  {Steffen}}]{Freytag96}
{Freytag}, B., {Ludwig}, H., \& {Steffen}, M. 1996, A\&A, {\bf 313}, 497

\bibitem[{{Fujimoto} \& {Murai}(1984)}]{1984IAUS..108..115F}
{Fujimoto}, M. \& {Murai}, T. 1984, in IAU Symp. 108: Structure and Evolution
  of the Magellanic Clouds, ed. S.~{van den Bergh} \& K.~S.~D. {Boer}, 115--123

\bibitem[{{Gardiner} \& {Hatzidimitriou}(1992)}]{gardiner1992}
{Gardiner}, L.~T. \& {Hatzidimitriou}, D. 1992, \mnras, 257, 195

\bibitem[{{Gardiner} {et~al.}(1994){Gardiner}, {Sawa}, \&
  {Fujimoto}}]{gardiner1994}
{Gardiner}, L.~T., {Sawa}, T., \& {Fujimoto}, M. 1994, \mnras, 266, 567

\bibitem[{{Geha} {et~al.}(1998){Geha}, {Holtzman}, {Mould}, {Gallagher},
  {Watson}, {Cole}, {Grillmair}, {Stapelfeldt}, {Ballester}, {Burrows},
  {Clarke}, {Crisp}, {Evans}, {Griffiths}, {Hester}, {Scowen}, {Trauger}, \&
  {Westphal}}]{geha1998}
{Geha}, M.~C., {Holtzman}, J.~A., {Mould}, J.~R., {et~al.} 1998, \aj, 115, 1045

\bibitem[{{Girardi} {et~al.}(2002){Girardi}, {Bertelli}, {Bressan}, {Chiosi},
  {Groenewegen}, {Marigo}, {Salasnich}, \& {Weiss}}]{girardi2002}
{Girardi}, L., {Bertelli}, G., {Bressan}, A., {et~al.} 2002, \aap, 391, 195

\bibitem[{{Girardi} {et~al.}(2000){Girardi}, {Bressan}, {Bertelli}, \&
  {Chiosi}}]{girardi2000}
{Girardi}, L., {Bressan}, A., {Bertelli}, G., \& {Chiosi}, C. 2000, \aaps, 141,
  371

\bibitem[{{Harris} \& {Zaritsky}(1999)}]{harris1999}
{Harris}, J. \& {Zaritsky}, D. 1999, \aj, 117, 2831

\bibitem[{{Harris} \& {Zaritsky}(2004)}]{harris2004}
{Harris}, J. \& {Zaritsky}, D. 2004, \aj, 127, 1531

\bibitem[{{Hill} \& {Zaritsky}(2006)}]{hill2006}
{Hill}, A. \& {Zaritsky}, D. 2006, \aj, 131, 414

\bibitem[{{Hodge}(1986)}]{hodge1986}
{Hodge}, P. 1986, \pasp, 98, 1113

\bibitem[{{Hunter} {et~al.}(2003){Hunter}, {Elmegreen}, {Dupuy}, \&
  {Mortonson}}]{hunter2003}
{Hunter}, D.~A., {Elmegreen}, B.~G., {Dupuy}, T.~J., \& {Mortonson}, M. 2003,
  \aj, 126, 1836

\bibitem[{{Javiel} {et~al.}(2005){Javiel}, {Santiago}, \&
  {Kerber}}]{javiel2005}
{Javiel}, S.~C., {Santiago}, B.~X., \& {Kerber}, L.~O. 2005, \aap, 431, 73

\bibitem[{{Kerber} \& {Santiago}(2006)}]{2006A&A...452..155K}
{Kerber}, L.~O. \& {Santiago}, B.~X. 2006, \aap, 452, 155

\bibitem[{{King}(1962)}]{king1962}
{King}, I. 1962, \aj, 67, 471

\bibitem[{{Kroupa}(2000)}]{kroupa2000}
{Kroupa}, P. 2000, in Astronomische Gesellschaft Meeting Abstracts, ed. R.~E.
  {Schielicke}, 11

\bibitem[{{Lattanzio} {et~al.}(1991){Lattanzio}, {Vallenari}, {Bertelli}, \&
  {Chiosi}}]{L91}
{Lattanzio}, J., {Vallenari}, A., {Bertelli}, G., \& {Chiosi}, C. 1991, A\&A,
  {\bf 250}, 340

\bibitem[{{Lisenfeld} \& {Ferrara}(1998)}]{lisenfeld1998}
{Lisenfeld}, U. \& {Ferrara}, A. 1998, \apj, 496, 145

\bibitem[{{Mackey} \& {Gilmore}(2003)}]{2003MNRAS.338..120M}
{Mackey}, A.~D. \& {Gilmore}, G.~F. 2003, \mnras, 338, 120

\bibitem[{{Maeder}(1975)}]{maeder1975}
{Maeder}, A. 1975, \aap, 43, 61

\bibitem[{{Mathewson} {et~al.}(1986){Mathewson}, {Ford}, \&
  {Visvanathan}}]{mathewson1986}
{Mathewson}, D.~S., {Ford}, V.~L., \& {Visvanathan}, N. 1986, \apj, 301, 664

\bibitem[{{McLaughlin} \& {van der Marel}(2005)}]{2005ApJS..161..304M}
{McLaughlin}, D.~E. \& {van der Marel}, R.~P. 2005, \apjs, 161, 304

\bibitem[{{Mermilliod} \& {Mayor}(1989)}]{Mermi-Mayor89}
{Mermilliod}, J. \& {Mayor}, M. 1989, A\&A, {\bf 219}, 15

\bibitem[{{Noel} {et~al.}(2006){Noel}, {Gallart}, {Costa}, \&
  {Mendez}}]{Noel2006}
{Noel}, N., {Gallart}, C., {Costa}, E., \& {Mendez}, R.~A. 2006, Rev. Mex.
  Astronomia y Astrofizica, 26, 76

\bibitem[{{Olsen}(1999)}]{olsen1999}
{Olsen}, K.~A.~G. 1999, \aj, 117, 2244

\bibitem[{{Olszewski} {et~al.}(1996){Olszewski}, {Suntzeff}, \&
  {Mateo}}]{olszewski1996}
{Olszewski}, E.~W., {Suntzeff}, N.~B., \& {Mateo}, M. 1996, \araa, 34, 511

\bibitem[{{Pagel} \& {Tautvai{\v s}ien{\. e}}(1999)}]{pagel1999}
{Pagel}, B.~E.~J. \& {Tautvai{\v s}ien{\. e}}, G. 1999, \apss, 265, 461

\bibitem[{{Pasetto} {et~al.}(2003){Pasetto}, {Chiosi}, \&
  {Carraro}}]{pasetto2003}
{Pasetto}, S., {Chiosi}, C., \& {Carraro}, G. 2003, A\&A, 405, 931

\bibitem[{{Piatti} {et~al.}(2001){Piatti}, {Santos}, {Clari{\' a}}, {Bica},
  {Sarajedini}, \& {Geisler}}]{piatti2001}
{Piatti}, A.~E., {Santos}, J.~F.~C., {Clari{\' a}}, J.~J., {et~al.} 2001,
  \mnras, 325, 792

\bibitem[{{Pietrinferni} {et~al.}(2004){Pietrinferni}, {Cassisi}, {Salaris}, \&
  {Castelli}}]{pietrinferni2004}
{Pietrinferni}, A., {Cassisi}, S., {Salaris}, M., \& {Castelli}, F. 2004, \apj,
  612, 168

\bibitem[{{Pietrzynski} \& {Udalski}(1999)}]{pietrzynski1999}
{Pietrzynski}, G. \& {Udalski}, A. 1999, Acta Astronomica, 49, 157

\bibitem[{{Prather} \& {Demarque}(1974)}]{prather1974}
{Prather}, M.~J. \& {Demarque}, P. 1974, \apj, 193, 109

\bibitem[{{Rachford} {et~al.}(2002){Rachford}, {Snow}, {Tumlinson}, {Shull},
  {Blair}, {Ferlet}, {Friedman}, {Gry}, {Jenkins}, {Morton}, {Savage},
  {Sonnentrucker}, {Vidal-Madjar}, {Welty}, \& {York}}]{tumlinson2002}
{Rachford}, B.~L., {Snow}, T.~P., {Tumlinson}, J., {et~al.} 2002, \apj, 577,
  221

\bibitem[{{Rafelski} \& {Zaritsky}(2005)}]{rafelski2005}
{Rafelski}, M. \& {Zaritsky}, D. 2005, \aj, 129, 2701

\bibitem[{{Rich} {et~al.}(2000){Rich}, {Shara}, {Fall}, \& {Zurek}}]{rich2000}
{Rich}, R.~M., {Shara}, M., {Fall}, S.~M., \& {Zurek}, D. 2000, \aj, 119, 197

\bibitem[{{Riess} \& {Mack}(2004)}]{riess2004}
{Riess}, A. \& {Mack}, J. 2004, ISR ACS, 2004

\bibitem[{{Salpeter}(1955)}]{1955ApJ...121..161S}
{Salpeter}, E.~E. 1955, \apj, 121, 161

\bibitem[{{Sandage} {et~al.}(1999){Sandage}, {Bell}, \&
  {Tripicco}}]{sandage1999}
{Sandage}, A., {Bell}, R.~A., \& {Tripicco}, M.~J. 1999, \apj, 522, 250

\bibitem[{{Shaviv} \& {Salpeter}(1973)}]{shaviv1973}
{Shaviv}, G. \& {Salpeter}, E.~E. 1973, \apj, 184, 191

\bibitem[{{Sirianni} {et~al.}(2005){Sirianni}, {Jee}, {Ben{\'{\i}}tez},
  {Blakeslee}, {Martel}, {Meurer}, {Clampin}, {De Marchi}, {Ford}, {Gilliland},
  {Hartig}, {Illingworth}, {Mack}, \& {McCann}}]{sirianni2005}
{Sirianni}, M., {Jee}, M.~J., {Ben{\'{\i}}tez}, N., {et~al.} 2005, \pasp, 117,
  1049

\bibitem[{{Smecker-Hane} {et~al.}(2002){Smecker-Hane}, {Cole}, {Gallagher}, \&
  {Stetson}}]{smecker-hane2002}
{Smecker-Hane}, T.~A., {Cole}, A.~A., {Gallagher}, J.~S., \& {Stetson}, P.~B.
  2002, \apj, 566, 239

\bibitem[{{Stanimirovi{\' c}} {et~al.}(2004){Stanimirovi{\' c}},
  {Staveley-Smith}, \& {Jones}}]{stanimirovic2004}
{Stanimirovi{\' c}}, S., {Staveley-Smith}, L., \& {Jones}, P.~A. 2004, \apj,
  604, 176

\bibitem[{{Stanimirovic} {et~al.}(1999){Stanimirovic}, {Staveley-Smith},
  {Dickey}, {Sault}, \& {Snowden}}]{stanimirovic1999}
{Stanimirovic}, S., {Staveley-Smith}, L., {Dickey}, J.~M., {Sault}, R.~J., \&
  {Snowden}, S.~L. 1999, \mnras, 302, 417

\bibitem[{{Staveley-Smith} {et~al.}(1997){Staveley-Smith}, {Sault},
  {Hatzidimitriou}, {Kesteven}, \& {McConnell}}]{staveley1997}
{Staveley-Smith}, L., {Sault}, R.~J., {Hatzidimitriou}, D., {Kesteven}, M.~J.,
  \& {McConnell}, D. 1997, \mnras, 289, 225

\bibitem[{{Stetson}(1994)}]{stetson1994}
{Stetson}, P.~B. 1994, in The Restoration of HST Images and Spectra - II, ed.
  R.~J. {Hanisch} \& R.~L. {White}, 308

\bibitem[{{Storm} {et~al.}(2004){Storm}, {Carney}, {Gieren}, {Fouqu{\' e}},
  {Latham}, \& {Fry}}]{storm2004}
{Storm}, J., {Carney}, B.~W., {Gieren}, W.~P., {et~al.} 2004, \aap, 415, 531

\bibitem[{{Stothers} \& {Chin}(1981)}]{stothers1981}
{Stothers}, R. \& {Chin}, C.-W. 1981, \apj, 247, 1063

\bibitem[{{Stryker} {et~al.}(1985){Stryker}, {Da Costa}, \&
  {Mould}}]{stryker1985}
{Stryker}, L.~L., {Da Costa}, G.~S., \& {Mould}, J.~R. 1985, ApJ, 298, 544

\bibitem[{{Testa} {et~al.}(1999){Testa}, {Ferraro}, {Chieffi}, {Straniero},
  {Limongi}, \& {Fusi Pecci}}]{Te99}
{Testa}, V., {Ferraro}, F., {Chieffi}, A., {et~al.} 1999, AJ, {\bf 118}, 2839

\bibitem[{{Udalski} {et~al.}(1998){Udalski}, {Szymanski}, {Kubiak},
  {Pietrzynski}, {Wozniak}, \& {Zebrun}}]{Udalski1998}
{Udalski}, A., {Szymanski}, M., {Kubiak}, M., {et~al.} 1998, Acta Astronomica,
  48, 147

\bibitem[{{Unno} \& {Kondo}(1989)}]{UnnoKondo89}
{Unno}, W. \& {Kondo}, M. 1989, PASJ, {\bf 41}, 197

\bibitem[{{Vallenari} {et~al.}(1992){Vallenari}, {Chiosi}, {Bertelli},
  {Meylan}, \& {Ortolani}}]{VaC92}
{Vallenari}, A., {Chiosi}, C., {Bertelli}, G., {Meylan}, G., \& {Ortolani}, S.
  1992, AJ, {\bf 104}, 1100

\bibitem[{{Vallenari} {et~al.}(1991){Vallenari}, {Chiosi}, {Bertelli},
  {Meylan}, \& {Ortolani}}]{V91}
{Vallenari}, A., {Chiosi}, C., {Bertelli}, G., {Meylan}, S., \& {Ortolani}, S.
  1991, A\&AS, {\bf 87}, 517

\bibitem[{{Vallenari} {et~al.}(1996){Vallenari}, {Chiosi}, {Bertelli}, \&
  {Ortolani}}]{vallenari1996}
{Vallenari}, A., {Chiosi}, C., {Bertelli}, G., \& {Ortolani}, S. 1996, \aap,
  309, 358

\bibitem[{{Welch} {et~al.}(1987){Welch}, {McLaren}, {Madore}, \&
  {McAlary}}]{welch1987}
{Welch}, D.~L., {McLaren}, R.~A., {Madore}, B.~F., \& {McAlary}, C.~W. 1987,
  \apj, 321, 162

\bibitem[{{Weldrake} {et~al.}(2004){Weldrake}, {Sackett}, {Bridges}, \&
  {Freeman}}]{weldrake2004}
{Weldrake}, D.~T.~F., {Sackett}, P.~D., {Bridges}, T.~J., \& {Freeman}, K.~C.
  2004, \aj, 128, 736

\bibitem[{{Wilkinson} {et~al.}(2003){Wilkinson}, {Hurley}, {Mackey}, {Gilmore},
  \& {Tout}}]{2003MNRAS.343.1025W}
{Wilkinson}, M.~I., {Hurley}, J.~R., {Mackey}, A.~D., {Gilmore}, G.~F., \&
  {Tout}, C.~A. 2003, \mnras, 343, 1025

\bibitem[{{Wilson}(1975)}]{1975AJ.....80..175W}
{Wilson}, C.~P. 1975, \aj, 80, 175

\bibitem[{{Wolfire} {et~al.}(1995){Wolfire}, {Hollenbach}, {McKee}, {Tielens},
  \& {Bakes}}]{wolfire1995}
{Wolfire}, M.~G., {Hollenbach}, D., {McKee}, C.~F., {Tielens}, A.~G.~G.~M., \&
  {Bakes}, E.~L.~O. 1995, \apj, 443, 152

\bibitem[{{Xiong}(1980)}]{Xiong80}
{Xiong}, D.~R. 1980, ChA, {\bf 4}, 234

\bibitem[{{Yoshizawa} \& {Noguchi}(2003)}]{yoshizawa2003}
{Yoshizawa}, A.~M. \& {Noguchi}, M. 2003, \mnras, 339, 1135

\bibitem[{{Zahn}(1991)}]{Zahn91}
{Zahn}, J. 1991, A\&A, {\bf 252}, 179

\bibitem[{{Zaritsky} {et~al.}(2002){Zaritsky}, {Harris}, {Thompson}, {Grebel},
  \& {Massey}}]{zaritsky2002}
{Zaritsky}, D., {Harris}, J., {Thompson}, I.~B., {Grebel}, E.~K., \& {Massey},
  P. 2002, \aj, 123, 855

\end{thebibliography}

\end{document}